# High-pressure electride superconductor $Li_5N$ for multifunctional applications: A theoretical insight into the physical properties

M. Abdul Hadi Shah[a,b], S.H. Naqib[b,*]

[a]Department of Physics, Rajshahi University of Engineering and Technology, Rajshahi 6204, Bangladesh
[b]Department of Physics, University of Rajshahi, Rajshahi 6205, Bangladesh
*Corresponding author: Email: salehnaqib@yahoo.com

**Abstract**

This study aims to unveil the physical properties of multifunctional $Li_5N$ electride under high pressure in the range of 150-350 GPa through first-principles analysis within the density functional theory (DFT). The calculated structural parameters are in fair agreement with the small number of data available. The estimated negative formation energy indicates that $Li_5N$ is chemically stable and synthesizable within the studied pressure range. A thorough analysis of phonon dispersion was conducted from 0 GPa to 400 GPa, revealing dynamic stability between 100 GPa and 382 GPa. The pressure dependence of the elastic stiffness constants together with their mechanical and thermo-physical features is investigated in-depth. The compound is categorized as ductile in light of estimated Poisson's ratio, Pugh's ratio and Cauchy pressure. Electronic band structure and density of states analyses reveal metallic feature for all pressures. The density of states at Fermi level diminishes with increasing pressure. The superconducting state properties of $Li_5N$ under pressure are also discussed. The optical response of $Li_5N$ clearly demonstrates significant intraband contributions at low energies, robust ultraviolet absorption (~ $10^5$ $cm^{-1}$), and impressive low-energy reflectivity. The systematic variation of all structural, mechanical, optoelectronic, and superconducting properties across the studied pressure range manifests an effective computational strategy for tuning multifunctional properties in $Li_5N$ electride.

## 1. Introduction

Electrides are a distinctive class of ionic solids with a potential chemical curiosity in which electrons act as anions [1,2]. Unlike the nearly free and highly delocalized electrons in conventional metals, the excess electrons in electrides are confined within the interstitial cavities (where they act as anions). These interstitial electrons exhibit pronounced quantum effects owing to the exceptionally small mass of electrons compared with conventional anions. Moreover, their weak binding to the surrounding cationic framework endows electrides with remarkable physical and chemical properties, including high electron conductivity [3], low work function [4], strong nonlinear optical responses, namely, hyperpolarizabilities [5], and excellent catalytic activity [6].

Since the discovery of the first crystalline organic electride (Cesium 18-Crown-6 Compounds) in 1983 synthesized by Dye *et al.*, [7], a number of organic electrides have been synthesized and their crystal structures were explored [8–11]. Although organic electrides are good low-temperature electron emitters and strong reducing agents [12,13], they are often thermally unstable and may degrade quickly with increasing temperatures. In addition, sensitivity to oxygen and moisture continues to pose a significant challenge in the synthesis of organic electrides [14]. Addressing these challenges might significantly confine their potential applications, and materials scientists are now decisively focusing on inorganic electrides [15–19]. The first room-temperature, air-stable inorganic electride was introduced by S. Matsuishi *et al.* in 2003 [3]. This compound possesses a high electron density with low work function [3]. Therefore, they are promising for next-generation electronics, making them suitable for display devices [20], and used as efficient cathode material with low electron-injection barrier in organic light-emitting diodes (OLED) [21].

Depending on the topologies and dimensionality of interstitial electrons, electrides are generally classified into zero-dimensional (0D), one-dimensional (1D), two-dimensional (2D), and three-dimensional (3D). The anionic electrons are trapped in atomic cages in 0D electrides in which the highly localized anionic electrons lead to a narrow band within the band structure, named as cage states [3,17]. In 1D electrides, the anionic electrons are distributed in the channel voids, forming 1D electron gas [22]. The anionic electrons in 2D electrides [2,23] demonstrate higher dimensionality and exhibit more delocalized energy bands compared to 0D and 1D [24]. In contrast, compounds such as $Li_4N$ [25], $Li_5N$ [26], $Li_5C$ [27], $Nb_5Ir_3N$ [28], $Ca_2C$ [13], and several subnitrides [29] have been theoretically proposed as 3D electrides, their direct experimental verification of 3D interconnected interstitial electron networks remains limited. Consequently, experimentally established 3D electrides are still scarce.

Despite the discovery of many electrides at ambient pressure [30,31], external compression remains a powerful and often indispensable approach for stabilizing unconventional electride states and uncovering emergent phenomena, including superconductivity, metal–insulator transitions, and exotic electronic behavior [32]. Therefore, electrides constitute a unique class of materials in which high pressure plays a crucial role in stabilizing novel electride phases and tailoring their electronic properties, attracting considerable research interest over the past few decades [26,32–38].

The coexistence of superconductivity and electride states represents an exciting paradigm in condensed matter physics, offering unprecedented opportunities for designing novel quantum materials with exotic physical properties [39]. However, the pursuit of high-transition temperature (high-$T_c$) superconductivity stands as a leading frontier in condensed matter physics research [40]. Inorganic electrides are not generally recognized as a distinct class of high-$T_c$ superconductors at ambient conditions. Rather, several inorganic electrides have been found or predicted to exhibit superconductivity, which has stimulated interest in exploring electrides as potential superconducting materials. Electrides at ambient pressure are known to exhibit weak superconducting features. For instance, synthesized $Y_2C$ [40], $Nb_5Ir_3$ [39], and $Nb_5Ir_3N$ [28] all exhibit critical temperature ($T_c$) below 10 K. Addressing this, high pressure has emerged as a powerful approach for discovering high-$T_c$ superconducting electride, as it can stabilize electride phases that are inaccessible at ambient pressure, and substantially modify their electronic and superconducting properties.

High-pressure electrides (HPEs) often composed of alkaline earth and non-metallic elements [42]. A number of research have been conducted on Li-C, Li-P, and Li-Si binary systems with the isomorphic symmetry as Li-N [27,36–38]. For example, Z. Zhao *et al.* investigated stable high pressure phase $Li_5P$ [43] as a promising anode material for Li-ion batteries, while Z. Zhao *et al.* [40] predicted $Li_6P$ electride with estimated $T_c$ of 39.3 K at 270 GPa. In contrast, Z.S. Pereira *et al.* predicted 2D electride $Li_5C$ at high pressures in the range from 50 to 210 GPa with hexagonal symmetry ($P6/mmm$) [27], estimated $T_c$ of 48.3 K at 210 GPa (highest $T_c$ among all predicted Li-C compounds so far), and one of the highest known $T_c$ predicted for electrides (exceeds $Li_6P$).

Nowadays, Li-N systems are investigated with immense interest due to their potential uses as electrolytes in Li-ion batteries and energy storage areas [44–46]. However, only a limited number of studies have been conducted on Li-N systems with and without electride features under ambient and high pressures [33,37,47,48]. Among them, five novel stable Li-N compounds ($Li_{13}N$, $Li_5N$, $Li_3N_2$, $LiN_2$, and $LiN_5$) within 0 to 100 GPa pressure are predicted via *ab initio* simulations [37]. They identified $Li_5N$ as a thermodynamically stable structure over the pressure range of 80 GPa to at least 100 GPa, adopting a single $P6/mmm$ crystal structure throughout the interval. Remarkably, their study neither explored the possibility of electride behavior nor investigated superconductivity, thereby leaving these intriguing properties unexplored and motivating future investigations, particularly at pressures beyond 100 GPa.

Later on, Z. Wan *et al.* was investigated $Li_5N$ electride as hexagonal anionic electron topology under high pressure within 150 GPa to 350 GPa [26]. Due to exhibiting multifunctional features, $Li_5N$ becoming the promising candidate to be investigated intensively. Although the key superconducting features are conducted at high pressures, many of the physical properties such as structural, phonon, elastic, thermophysical, anisotropic, electronic, and optical behavior still remain to be unveiled. Therefore, this study intends to explore the fundamental physical nature of $Li_5N$ under pressure within the pressure range between 150 GPa – 350 GPa.

## 2. Computational methodology

### *2.1. Structural and electronic calculations*

Employing plane-wave pseudopotential based approach, DFT [49] simulation within the first-principles investigation as integrated in the CASTEP package [50] have been utilized in this work to explore structural and electronic features of $Li_5N$ electride. Perdew-Burke-Ernzerhof with solids-corrected (PBEsol) [51] in GGA is chosen for exchange-correlation functional. An ultrasoft pseudopotential of Vanderbilt-type is adopted for the interaction of charges between atomic cores and valence electrons. The solids-corrected version of the GGA approximation developed by Perdew *et al*. for densely packed solids offers a precise insight into their physical nature than the conventional version [52]. A plane-wave basis set cut-off of 500 eV and Monkhorst-Pack $k$-grid of 14×14×9 (120 irreducible $k$-points) was taken in the Brillouin zone (BZ) integrations for achieving suitable convergence. Broyden–Fletcher–Goldfarb–Shanno (BFGS)-type algorithm [53] is used to optimize equilibrium structural parameter. A self-consistent field (SCF) tolerance of $2.0\times10^{-6}$ eV/atom was set as a difference in the total energy. Pulay density mixing scheme along with a charge mixing amplitude of 0.5 and charge mixing cut-off of 1.5 was selected for the electronic minimizer. Other essential convergence parameters during optimization process are chosen as: energy of the system: $1.0\times10^{-5}$ eV/atom, interaction force between the atoms: 0.03 eV/Å, stress: 0.05 GPa and maximum displacement: 0.001 Å.

Phonon dispersion curves (PDC) and phonon density of states (PHDOS) are calculated within the density functional perturbation theory (DFPT) based on the finite displacement supercell method [54,55]. A supercell volume of eight times that of the unit cell of $Li_5N$ and a cutoff radius of 2.0 Å were employed during phonon calculations keeping energy cut-off and $k$-grid similar for all pressures.

*2.2. Numerical details*

The elastic constants of materials can be estimated by the stress-strain method [56] implemented within the Hooke's law, $\sigma_{ij} = C_{ijkl}\varepsilon_{kl}$, where $\sigma_{ij}$ is the stress tensor, $C_{ijkl}$ is the elastic constant tensor which is a 6×6 matrix, and $\varepsilon_{kl}$ represents the Lagrangian strain tensor.

The diversity of symmetry of different crystal structure could result in some tensors equal and others equal to zero. There are twenty one independent elastic constants $C_{ij}$ as a whole, but the structural symmetry of hexagonal crystals, the matrix of stiffness constants can be expressed in terms of five independent constants ($C_{11}$, $C_{12}$, $C_{13}$, $C_{33}$, and $C_{44}$) [57] and one dependent constant ($C_{66}$) [58]:

$$C = \begin{pmatrix} C_{11} & C_{12} & C_{13} & 0 & 0 & 0 \\ C_{12} & C_{11} & C_{13} & 0 & 0 & 0 \\ C_{13} & C_{13} & C_{33} & 0 & 0 & 0 \\ 0 & 0 & 0 & C_{44} & 0 & 0 \\ 0 & 0 & 0 & 0 & C_{44} & 0 \\ 0 & 0 & 0 & 0 & 0 & \frac{C_{11}-C_{12}}{2} \end{pmatrix}. \quad (1)$$

Single-crystal elastic constants allow us to estimate the polycrystalline features within shear modulus $G$ and bulk modulus $B$ following Voigt–Reuss–Hill (VRH) approximation, which can be derived in terms of the independent elastic components as follows [59]:

$$G = \frac{G_V + G_R}{2}; \quad B = \frac{B_V + B_R}{2}, \quad (2)$$

where $B_R$ ($B_V$) and $G_R$ ($G_V$) represent the lower (upper) limit for polycrystalline crystal at the Reuss (Voigt) boundary.

The shear modulus for the Voigt ($G_V$) and Reuss ($G_R$) models, as well as the bulk modulus for Voigt ($B_V$) and the Reuss ($B_R$), are calculated for hexagonal lattices using following equations [60]:

$$\left.\begin{aligned} G_V &= \frac{1}{30}(C_{11} + C_{12} + 2C_{33} - 4C_{13} + 12C_{44} + 12C_{66}); \\ G_R &= \frac{5[(C_{11} + C_{12})C_{33} - 2C_{13}^2]C_{44}C_{66}}{6B_V C_{44}C_{66} + 2[(C_{11} + C_{12})C_{33} - 2C_{13}^2](C_{44} + C_{66})}; \\ B_V &= \frac{1}{9}[2(C_{11} + C_{12}) + 4C_{13} + C_{33}]; \\ B_R &= \frac{(C_{11} + C_{12})C_{33} - 2C_{13}^2}{C_{11} + C_{12} + 2C_{33} - 4C_{13}}. \end{aligned}\right\} \quad (3)$$

Moreover, the polycrystalline Young's modulus ($E$) and Poisson's ratio ($\sigma$) are computed as [61]:

$$E = \frac{9BG}{3B + G}; \ \sigma = \frac{3B - 2G}{2(3B + G)}. \quad (4)$$

The optical properties of materials are characterized by their dielectric function $\varepsilon(\omega)$ which is a complex tensor that explains the linear response of electronic system to the electromagnetic radiation.

$$\varepsilon(\omega) = \varepsilon_1(\omega) + \varepsilon_2(\omega), \quad (5)$$

where $\varepsilon_1(\omega)$ and $\varepsilon_2(\omega)$ corresponds to the real and imaginary parts of the dielectric constants, respectively. The frequency-dependent imaginary dielectric function $\varepsilon_2(\omega)$ indicate the absorption of the incident radiations and is expressed as [62]:

$$\varepsilon_2(\omega) = \left(\frac{e^2\hbar}{\pi m^2\omega^2}\right)\sum_{v,c}\int_{BZ} |M_{cv}(k)|^2\delta[\omega_{cv}(k) - \omega]d^3k, \quad (6)$$

where $M_{cv}(k) = \langle u_{ck}|\delta\nabla|u_{vk}\rangle$ describes the momentum dipole matrix components for direct transitions in between valence $u_{vk}(r)$ and conduction band $u_{ck}(r)$ within the wave vector $k$, $\nabla$ is the momentum operator, $\delta$ is the potential vector of the electric field, and $\hbar\omega_{cv}(k) = E_{ck} - E_{vk}$ corresponds to the transition energy. The integral is taken over the first Brillouin zone.

The real part of the dielectric constant, $\varepsilon_1(\omega)$, can be derived from the imaginary part $\varepsilon_2(\omega)$ using Kramers-Kronig relation as [63]:

$$\varepsilon_1(\omega) = 1 + \frac{2}{\pi}P\int_0^\infty \frac{\omega'\varepsilon_2(\omega')}{\omega'^2 - \omega^2}d\omega', \quad (7)$$

where $P$ signifies the principal value of the integral.

To acquire deeper understanding of optical nature, the complex refractive index $N(\omega)$ is to be estimated. Being a complex quantity, it has two parts: the real part, referred to as refractive index $n(\omega)$ and the imaginary part, known as the extinction coefficient, $k(\omega)$. Therefore, $N(\omega)$ is calculated as $N(\omega) = n(\omega) + ik(\omega)$; where $n(\omega)$ and $k(\omega)$ are estimated by using the expressions given in [64,65]. The other optical parameters *viz.* absorption coefficient $\alpha(\omega)$, optical conductivity $\sigma(\omega)$, reflectivity $R(\omega)$ and energy-loss function $L(\omega)$ are calculated employing the conventional approach [66].

## 3. Results and analysis

### *3.1. Structure and stability*

$Li_5N$, isomorphous to $Li_5P$, belongs to hexagonal structure having space group *P*6/*mmm* (#191) that contains one formula unit. In the structure, Li atoms occupy two nonequivalent positions; firstly, Li atoms in elementary cell are occupied at Li(1*a*): (0.000, 0.000, 0.000) and secondly, Li(4*h*): (0.333, 0.637, 0.290) positions, while N(1*b*): (0.000, 0.000, 0.500) is positioned such that each N atom has 14 nearest-neighbor Li atoms, forming a N−Li octadecahedron. Li (1*a*) symmetrically distribute on both sides of graphene-like layered structure along the *c*-axis while Li (4*h*) form graphene-like layered structures in the *ab* plane [26,43]. **Figure 1a-b** depicts the unit cell structure of $Li_5N$ in both two dimensions (2D: **Fig. 1a**) and three dimensions (3D: **Fig. 1b**), using the VESTA software.

In order to acquire equilibrium structural parameters at 150 GPa pressure, we have considered various functionals. Among those GGA-PBESol provides the least deviation of volume (~0.11%) regarding to the available data [26]; therefore, this functional is used to investigate the rest of the physical properties of $Li_5N$.

The variation of total energy with pressure and volume for $Li_5N$ electride is depicted in **Fig. 1c-d**. It is seen that the total energy of the unit cell escalates monotonically with pressure (**Fig. 1c**) and decreases with volume (**Fig. 1d**). These behaviors are well-allied with the literatures [67,68].

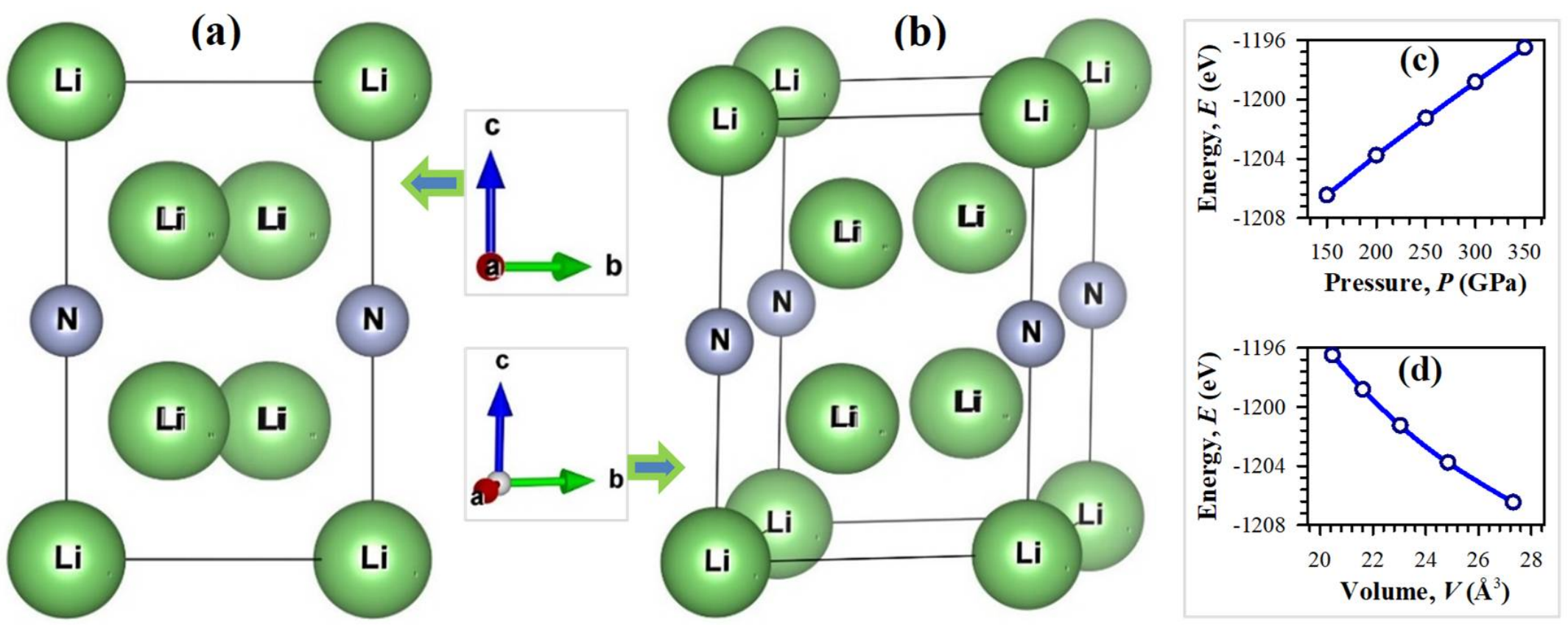

**Fig. 1**. Unit cell crystal structure of $Li_5N$ in real space: (a) two-dimensional view and (b) three-dimensional view, Figure on the right panel represents (c) pressure and (d) volume-dependent total energy of $Li_5N$.

Pressure-dependent lattice parameters ($a$, $c$), densities ($\rho$), and volume of the cell ($V$) for $Li_5N$ are illustrated in **Fig. 2**. The evaluated lattice constants and hexagonal ratio ($c/a$) (**Fig. 2a**), and their volume gradually decrease as pressure increases; meanwhile the densities increase (**Fig. 2b**). These findings indicate that as the pressure increases, the lattice parameter '$c$' decreases slightly faster than the lattice parameter '$a$'. Therefore, the lattice parameter '$c$' is more sensitive to external pressure than '$a$'. The whole trend are compatible with earlier literatures [67,68].

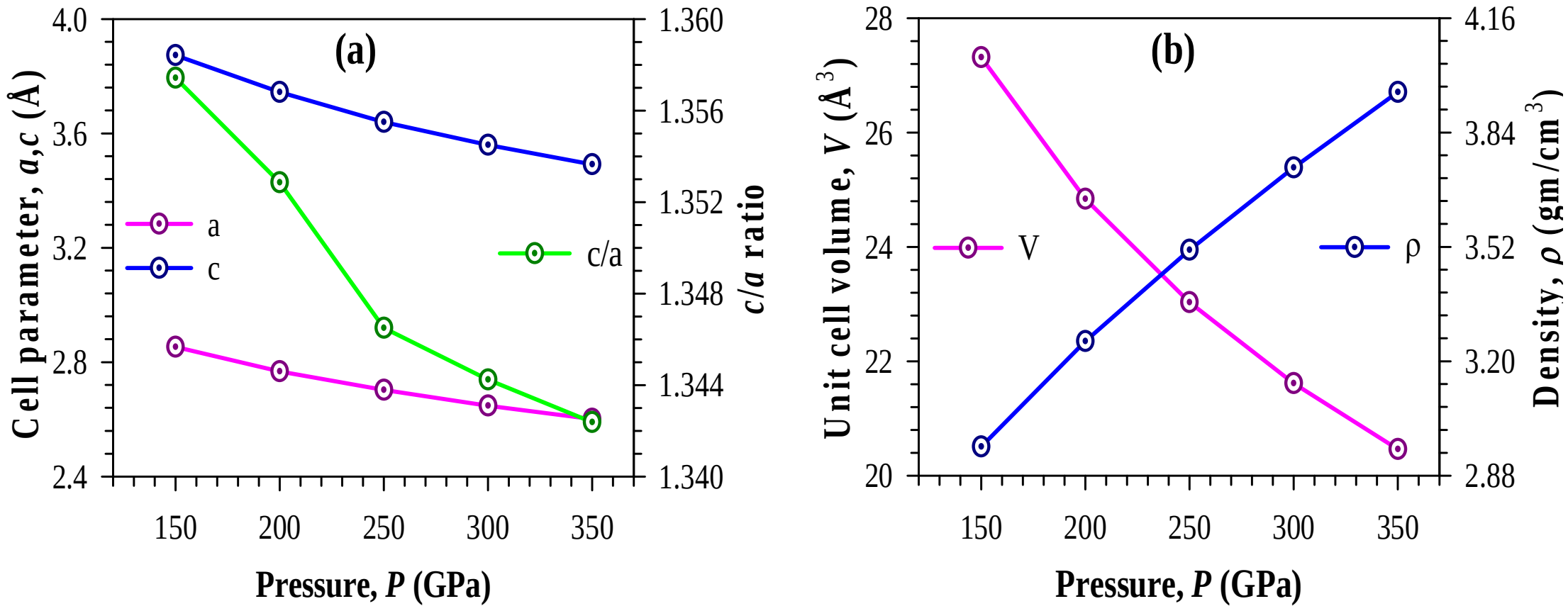

**Fig. 2**. Variation of (a) lattice parameter, (b) density ($\rho$), and volume ($V$) with pressure of $Li_5N$ electride.

**Table 1**. The equilibrium lattice constants ($a$, $b$, $c$) and related structural and energy parameters of $Li_5N$ under pressure.

| $P$ (GPa) | $a$ (Å) | $c$ (Å) | $c/a$ | $V$ (Å$^3$) | $E$ (eV) | $\rho$ (gm/cm$^3$) | $E_f$ (eV/atom) | $E_c$ (eV/atom) |
|---|---|---|---|---|---|---|---|---|
| 150 | 2.853 | 3.873 | 1.357 | 27.313 | -1206.470 | 2.961 | -0.176 | 1.671 |
| 200 | 2.768 | 3.744 | 1.353 | 24.839 | -1203.787 | 3.256 | -0.411 | 1.224 |
| 250 | 2.703 | 3.640 | 1.346 | 23.031 | -1201.277 | 3.511 | -0.600 | 0.806 |
| 300 | 2.648 | 3.559 | 1.344 | 21.613 | -1198.847 | 3.742 | -0.790 | 0.401 |
| 350 | 2.601 | 3.492 | 1.342 | 20.460 | -1196.513 | 3.953 | -0.962 | 0.012 |

### *3.1.1. Thermodynamic stability*

To validate the thermodynamic stability condition of $Li_5N$, its formation energy ($E_f$) and cohesive energy ($E_c$) are to be examined. The formation energy forecasts the stability of crystals in respect of decomposition into its bulk constituent elements; in contrary, the energy needed to form a crystal from free or isolated atoms is described as the cohesive energy. Both of them correlate with the structural stability and have been estimated for the given material using the relations [69,70]: $E_f = \frac{1}{6}\left(E_{Li_5N}^{total} - 5E_{Li}^{bulk} - E_{N}^{bulk}\right)$; $E_c = \frac{1}{6}\left(5E_{Li}^{iso} + E_{N}^{iso} - E_{Li_5N}^{total}\right)$, where $E_{Li_5N}^{\mathrm{total}}$ denotes the optimum energy of $Li_5N$, while $E_{\mathrm{Li}}^{\mathrm{bulk}}$ and $E_{\mathrm{N}}^{\mathrm{bulk}}$ reflect the total energies of the corresponding atom in which Li and N crystallizes in the bulk form in the BCC structure having space group ($Im\bar{3}m$, SG: 229) and ($Pa\bar{3}$, SG: 205), respectively. On the other hand, $E_{\mathrm{Li}}^{\mathrm{iso}}$ and $E_{\mathrm{N}}^{\mathrm{iso}}$ represent the energies of the corresponding isolated atoms. Generally, a negative value of $E_f$ and a positive values of $E_c$ indicate chemical stability [69,70].

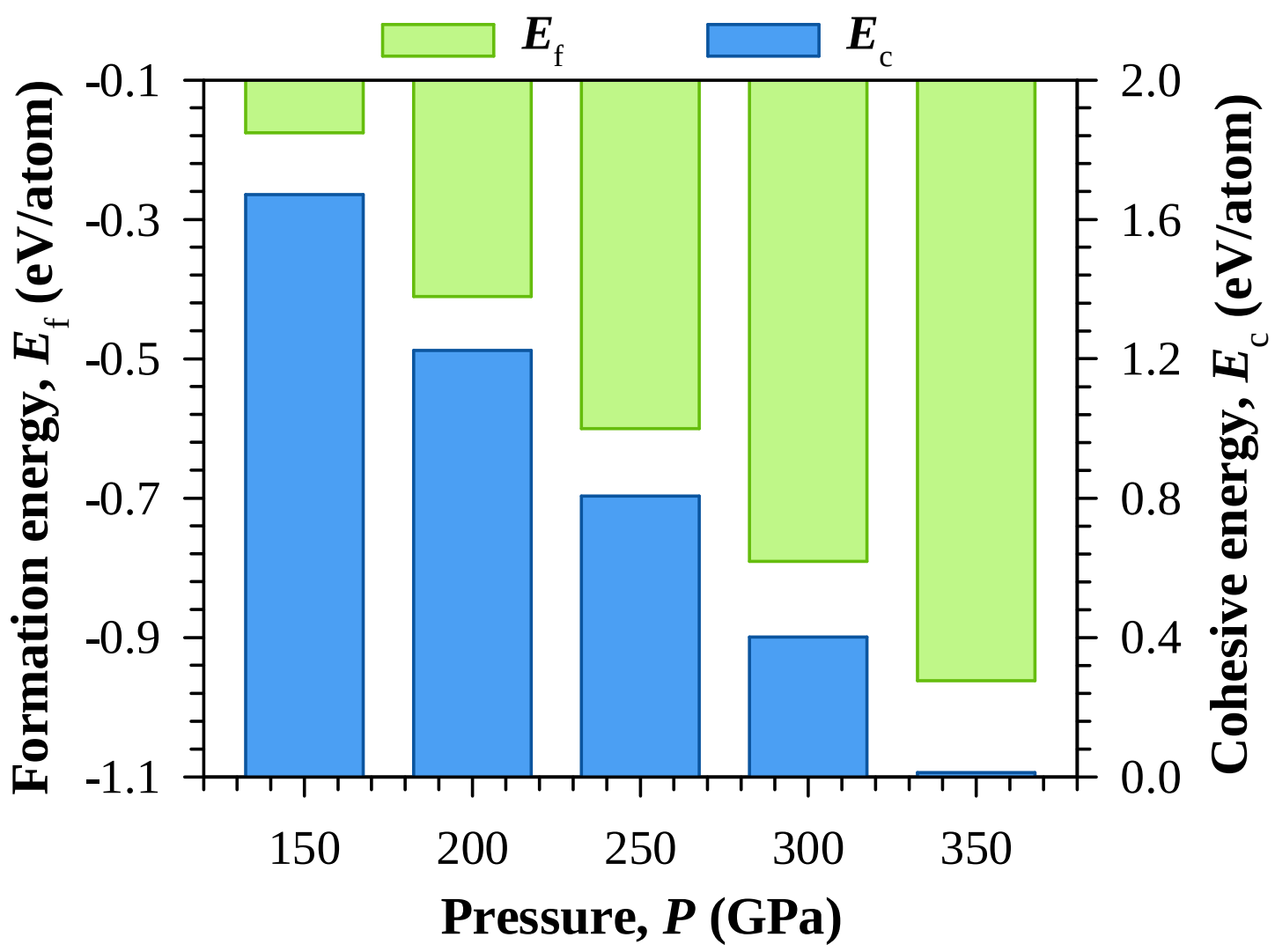

**Fig. 3**. Formation energy and cohesive energy as a function of pressure (150-350 GPa) for $Li_5N$.

The computed values of $E_f$ and $E_c$ (cf. **Table 1** and **Fig. 3**) are negative and positive, respectively, within the hydrostatic pressure 150-350 GPa implies that the studied $Li_5N$ in hexagonal symmetry are energetically feasible to synthesize from the thermodynamics point of view within the pressure range considered. It should be noted that a more negative $E_f$ value indicates better phase stability at 350 GPa. It is to be highlighted that the estimated formation energy ($E_f$) at 100 GPa is 0.107 eV/atom (positive), while the estimated cohesive energy ($E_c$) at 351 GPa is -0.002 eV/atom (negative), suggesting that the compound is in an unstable state at both 100 GPa and 351 GPa. The observed instability serves as a compelling rationale for further investigating $Li_5N$ within the pressure range of 150 to 350 GPa.

### *3.1.2. Dynamical stability and vibrational properties*

Phonon dispersion is an important phenomena of lattice dynamics that provide information regarding structural stability and vibrational contribution in the thermodynamic properties [71]. However, it is crucial to assess the structural stability of high-pressure phases in a compound [54]; consequently, we computed the phonon dispersion curves and phonon density of states (DOS) for different high-pressure phases of $Li_5N$. A compound is said to be dynamically stable when positive frequencies over the whole Brillouin zone appear while negative frequencies indicate the soft phonon modes and dynamical instability [54]. In this work, finite displacement supercell method within the first BZ is utilized in the CASTEP code to calculate the phonon dispersion curves from 0 GPa to 400 GPa hydrostatic pressures. In phonon dispersion calculations, the pressure is adjusted until a negative frequency is reached with a precision of 1 GPa. The calculated phonon dispersion curves of $Li_5N$ compound along the high symmetry directions are displayed in **Fig. 4a-i**. The zone center is set at the $\Gamma$ point, where the frequencies for the acoustic modes (indicated in violet) are zero in all cases.

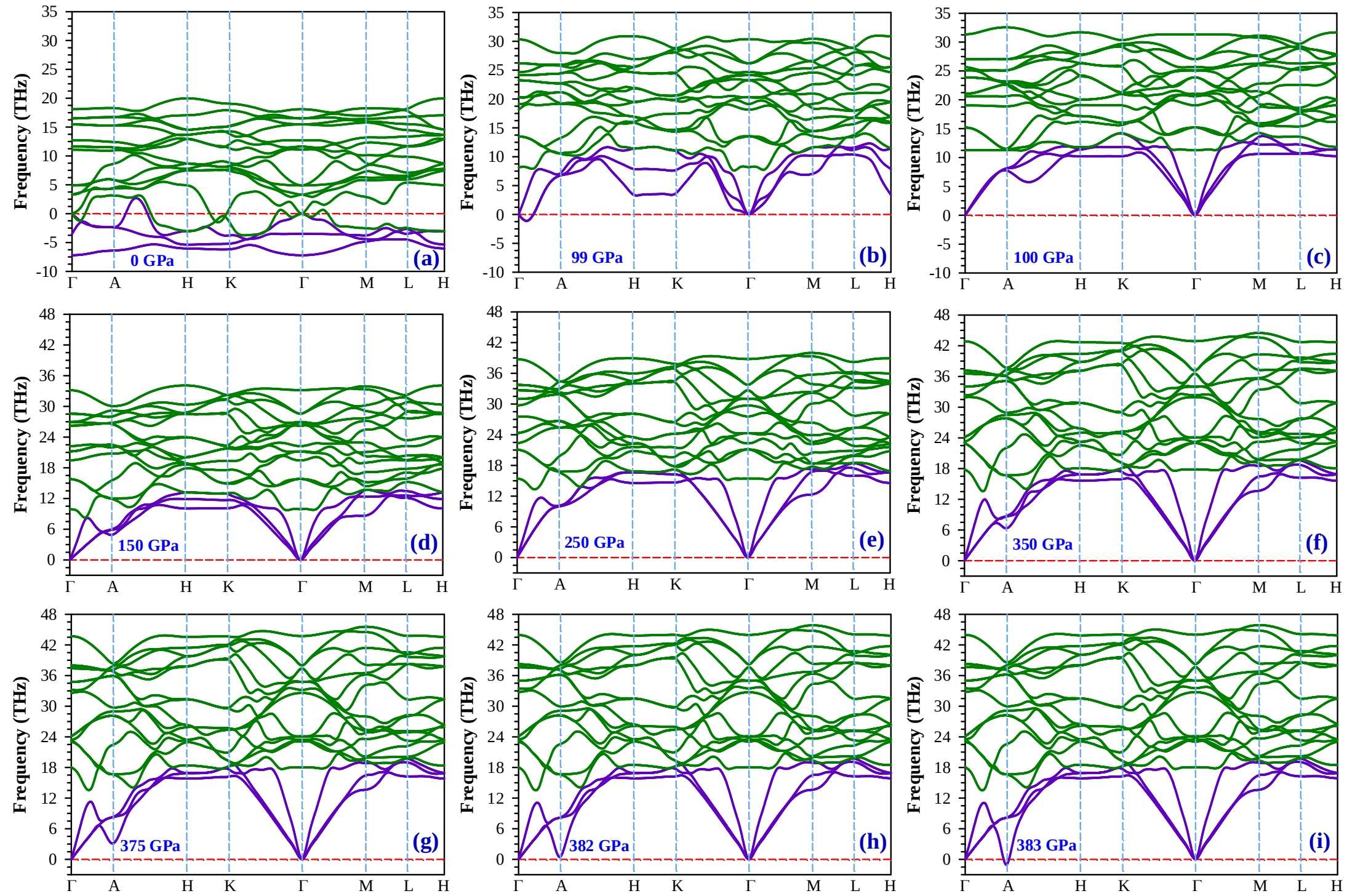

**Fig. 4**. Phonon dispersion spectra of $Li_5N$: (a) 0 GPa, (b) 99 GPa, (c) 100 GPa, (d) 150 GPa, (e) 250 GPa, (f) 350 GPa, (g) 375 GPa, (h) 382 GPa, and (i) 383 GPa.

Phonon dispersion consists of two types of modes: acoustic and optical. The acoustic modes consist of one longitudinal and two transverse acoustic branches, which arise from the coherent vibrations of atoms in a lattice as they move away from their equilibrium positions. In contrast, optical phonons originate from the out-of-phase oscillations of atoms in the lattice; when one atom moves to the left and its neighbor moves to the right [72,73]. For any material with '*n*' atoms per unit cell, there are three acoustic modes and ($3n$-3) optical modes. In the case of a unit cell of $Li_5N$, which contains six atoms, there are a total of eighteen vibrational phonon modes: three acoustic and fifteen optical modes for each *k*-mesh.

The absence of imaginary frequencies from 100 GPa to 382 GPa (**Fig. 4c-h**) indicates the phase stability from a dynamic point of view within this pressure range. However, from 0 to 99 GPa and above 383 GPa, instability is exhibited in $Li_5N$ due to the presence of negative frequencies (cf. **Fig. 4a-b** and **Fig. 4i**). This insight provides a valuable cornerstone for understanding the material's behavior under varying pressure conditions. The absence of energy gaps between the acoustic modes (lower branches) and the optical modes (upper branches indicated in dark green) highlights the consistent presence of elastic impedance across all studied pressures [68]. Phonon dispersion can also correlate bond lengths between atoms in compounds. The pressure-dependent increase in phonon dispersion modes correlates with a decrease in bond length, which is further supported by atomic population analysis (will be discussed in **section 3.6.2**).

**Table 2**. Calculated phonon frequencies at zone-center ($\Gamma$-point) of infrared (*IR*)- and Raman (*R*)-active optical modes at the of $Li_5N$ electride at various pressures.

| Symmetry | Types (number) of modes | Frequency ($cm^{-1}$) at | | | Active modes for | | | | | |
|---|---|---|---|---|---|---|---|---|---|---|
| | | | | | Infrared (*IR*) at | | | Raman (*R*) at | | |
| | | 150 GPa | 250 GPa | 350 GPa | 150 GPa | 250 GPa | 350 GPa | 150 GPa | 250 GPa | 350 GPa |
| $E_{1u}$ | Acoustic (2) | 0 | 0 | 0 | × | × | × | × | × | × |
| $A_{2u}$ | Acoustic (1) | 0 | 0 | 0 | × | × | × | × | × | × |
| $E_{1u}$ | Optical (2) | 706.55 | 810.23 | 804.09 | √ | √ | √ | × | × | × |
| $E_{1u}$ | Optical (2) | 898.87 | 1090.00 | 1219.27 | √ | √ | √ | × | × | × |
| $A_{2u}$ | Optical (1) | 741.56 | 919.51 | 1065.05 | √ | √ | √ | × | × | × |
| $A_{2u}$ | Optical (1) | 898.00 | 991.21 | 1078.68 | √ | √ | √ | × | × | × |
| $B_{2g}$ | Optical (1) | 330.16 | 515.23 | 595.26 | × | × | × | × | × | × |
| $E_{1g}$ | Optical (2) | 528.14 | 705.01 | 767.27 | × | × | × | √ | √ | √ |
| $B_{1u}$ | Optical (1) | 648.84 | 744.34 | 773.47 | × | × | × | × | × | × |
| $E_{2u}$ | Optical (2) | 874.91 | 1033.52 | 1134.18 | × | × | × | × | × | × |
| $E_{2g}$ | Optical (2) | 952.91 | 1125.53 | 1238.86 | × | × | × | √ | √ | √ |
| $A_{1g}$ | Optical (1) | 1107.06 | 1293.71 | 1430.69 | × | × | × | √ | √ | √ |

Based on the factor-group theory [55], irreducible representations of the vibrational modes in $Li_5N$ electride at the $\Gamma$-points are defined as: $\Gamma^{\text{acoustic}} = 2E_{1u} + A_{2u}$ and $\Gamma^{\text{optical}} = 4E_{1u}^{IR} + 2A_{2u}^{IR} + B_{2g}^{S} + 2E_{1g}^{R} + B_{1u}^{S} + 2E_{2u}^{S} + 2E_{2g}^{R} + A_{1g}^{R}$, where *IR* and *R* represent infrared active and Raman active mode frequencies, respectively, while *S* denotes the inactive mode often known as the silent mode. Among all these vibrational modes: the first three modes ($2E_{1u}$ and $A_{2u}$) are acoustic with zero frequencies at the $\Gamma$-point, and the other modes include five Raman- and six infrared-active optical modes. The remaining four modes ($B_{2g}$, $B_{1u}$, and $2E_{2u}$) are optically inactive. **Table 2** summarizes all the modes along with the phonon frequencies at the zone-center for $Li_5N$ at pressures 150 GPa, 250 GPa, and 350 GPa. Unfortunately, there is no experimental or theoretical data available regarding phonon frequencies; therefore, this study could be valuable for future experimental research in spectroscopy.

Phonon dispersion indicates that $Li_5N$ is dynamically stable at hydrostatic pressures ranging from 100 GPa to 382 GPa. However, its structural stability is only confirmed within the narrower pressure range of 150 GPa to 350 GPa.

### *3.2. Mechanical stability and properties*

The elastic constants are essential to explain mechanical resistance of crystalline solids to externally applied stresses. Particularly, they provide important insights into the stability, strength, and stiffness of materials, highlighting the need for precise methods in their *ab initio* estimations. By recognizing that forces and elastic constants are derived from the first and second derivatives of the potentials, we can undertake calculations that will constructively enhance the accuracy of force evaluations in solids. The calculated elastic stiffness constants ($C_{ij}$) along with polycrystalline elastic moduli ($B$, $G$, $E$) for $Li_5N$ electride at various pressures are listed in **Table 3**. For a stable hexagonal symmetry, all five elastic constants should satisfy the following necessary and sufficient conditions [69]: $C_{44} > 0$, $(C_{11} - C_{12}) > 0$, $(C_{11} +$ $2C_{12})C_{33} > 2C_{13}^2$. In contrast, the mechanical stability criteria under pressure are [74,75]: $\tilde{c}_{44} > 0, \tilde{c}_{11} > |\tilde{c}_{12}|, \tilde{c}_{33}(\tilde{c}_{11} + \tilde{c}_{12}) > 2\tilde{c}_{13}^2$; where $\tilde{c}_{ii} = C_{ii} - P\ (i = 1\sim4)$, $\tilde{c}_{12} = C_{12} + P, \tilde{c}_{13} = C_{13} + P$. Within the pressure interval of 150–350 GPa, the calculated elastic constants satisfy the required stability conditions, indicating that $Li_5N$ electride remains mechanically stable.

**Table 3**. Calculated single- and polycrystalline elastic constants ($C_{ij}$, *B*, *G*, *E* all in GPa), Poisson's ratio ($\sigma$) and Pugh's ratio (*G*/*B*), machinability index ($\mu_M$), Lamé's constants ($\lambda$ and $\mu$), Kleinmann parameter ($\xi$), and average Vicker's hardness ($H_V^{Avg}$ in GPa) for $Li_5N$ at various pressures.

| *P* (GPa) | $C_{11}$ | $C_{12}$ | $C_{13}$ | $C_{33}$ | $C_{44}$ | $C_{66}$ | *B* | *G* | *E* | $\sigma$ | *G*/*B* | $\mu_M$ | $\lambda$ | $\mu$ | $\xi$ | $H_V^{Avg}$ |
|---|---|---|---|---|---|---|---|---|---|---|---|---|---|---|---|---|
| 150 | 787.13 | 346.64 | 282.24 | 818.93 | 217.30 | 220.25 | 468.31 | 228.84 | 590.37 | 0.29 | 0.49 | 2.16 | 315.74 | 228.84 | 0.61 | 26.62 |
| 250 | 1154.72 | 568.56 | 477.75 | 1196.12 | 288.71 | 293.08 | 728.06 | 304.85 | 802.55 | 0.32 | 0.42 | 2.52 | 524.83 | 304.85 | 0.66 | 31.37 |
| 350 | 1491.80 | 790.21 | 678.52 | 1552.89 | 350.65 | 350.79 | 981.77 | 364.52 | 973.12 | 0.33 | 0.37 | 2.80 | 738.75 | 364.52 | 0.70 | 34.48 |

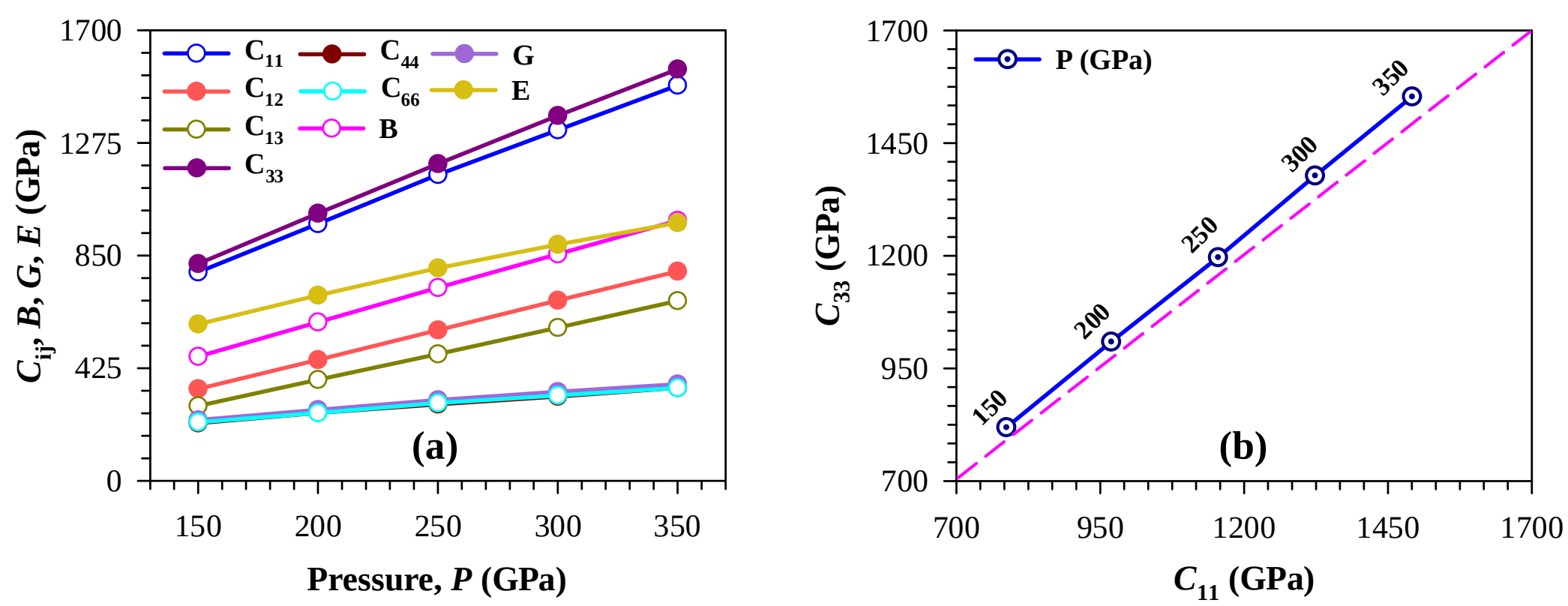

**Fig. 5**. (a) Elastic constants ($C_{ij}$) and moduli (*B*, *G*, *E*) as a function of hydrostatic pressure, and (b) $C_{33}$ versus $C_{11}$ at various pressures (150-350 GPa) for hexagonal $Li_5N$.

In general, the elastic constants $C_{11}$ and $C_{33}$ measure the resistance to linear compression along *a*- and *c*-directions, respectively. The calculated $C_{33}$ is higher than $C_{11}$ (**Table 3**), suggests that the compound $Li_5N$ is more compressible along *a*-axis than along *c*-axis [76]. This clearly aligns with the responses of both *a*- and *c*-axis when under pressure.

As seen in **Fig. 5a** and **Table 3**, the elastic constants $C_{11}$ and $C_{33}$ (which represent elasticity in length) are greater than the elastic constants $C_{12}$, $C_{13}$, $C_{44}$, and $C_{66}$ (which quantify elasticity in shape). This indicates that the resistance to deformation along the axial direction is stronger than the resistance to deformation in terms of shape of $Li_5N$. **Figure 5b** illustrates that as the values for $C_{33}$ increase, they demonstrate a proportional relationship with $C_{11}$. Moreover, the ratios of $C_{33}$ to $C_{11}$ are consistently greater than one, which suggests an intriguing distinction between the elastic responses along the *c*-axis compared to the *a*- and *b*-axes [77]. In contrast, the intrinsic hardness of a material is strongly correlated with $C_{44}$; the higher $C_{44}$ designates the higher hardness.

Polycrystalline nature of solids is characterized in light of bulk modulus (*B*), shear modulus (*G*), and Young's modulus (*E*). The bulk modulus measures how resistant a solid is to changes in volume, while the shear modulus measures material's resistance to shape change, and directly correlates with hardness. A larger shear modulus unequivocally indicates a higher $C_{44}$ value, reinforcing the relationship between these properties. The well-known Voigt–Reuss–Hill (VRH) approximation are used to estimate shear modulus $G$ and bulk modulus $B$ following **Eqs. 2-3**. The calculated bulk moduli and shear moduli indicate that at 350 GPa, $Li_5N$ is least compressible among all pressures (**Fig. 5a**). Moreover, the high bulk modulus with a low shear modulus of the compound clearly demonstrates the damage tolerant, easily machinable, quasi-ductile, and stiff nature of the materials [78]. Besides, Young's modulus for polycrystalline aggregates serves as a measure of the stiffness of solids; a large Young's modulus indicates stiffer materials. The studied $Li_5N$ is stiffest at 350 GPa for being the highest $E$. As the value of $E$ increases, the covalency of materials rises, peaking at 350 GPa, which indicates the highest ductility at this pressure as well [76].

In technological and engineering applications, investigating Poisson's ratio ($\nu$) is essential as it is a key elastic parameter that quantifies the stability of a crystal against shear forces. Generally, this ratio can take values between -1 and 0.5, which represents the lower limit ($\nu$ = -1), where the material does not change its shape and the upper limit ($\nu$ =0.5) when the volume remains unchanged. The calculated $\nu$ values, shown in **Fig. 6** and **Table 3**, fall within -1 and 0.5, indicating the stable linear elastic solids of $Li_5N$ under studied pressure [79]. The Poisson's ratio also estimates the ductility (or brittleness) and forecasts the degree of directionality of the bonding forces of materials. The value of $\nu$ greater than 0.26 makes materials ductile, while a value less than 0.26 makes them brittle, according to Frantsevich [80]. As pressure increases from 150 to 350 GPa, the value of $\nu$ for $Li_5N$ rises from 0.29 to 0.34. This indicates an improvement in ductility at higher pressures, which consequently enhances the plasticity [79]. Usually, a typical value of $\nu$ ~0.10 indicates the covalent bonding while the value of $\nu$ ~0.25 and ~0.33 corresponds to the presence of ionic and metallic bonding, respectively [81]. In view of calculated $\nu$, we conclude that the ionic and metallic contributions to the interatomic bonding are dominant in this structure. Moreover, the interatomic forces between the atoms are predominantly central in $Li_5N$ within the studied pressure range as the bonding forces meet the criteria for centrality when the value of $\nu$ is between 0.25 and 0.50 [61].

The Pugh index ($G/B$) is another measure used to assess the ductility (or brittleness) of materials [82]. A $G/B$ value of less (greater) than 0.57 indicates ductility (brittleness). As demonstrated in **Fig. 6**, all calculated $G/B$ values for this compound are below 0.57, clearly classifying it as ductile within the range of 150 to 350 GPa. Moreover, we observe that ductility and consequently plasticity significantly improves as pressure increases from 150 GPa to higher levels.

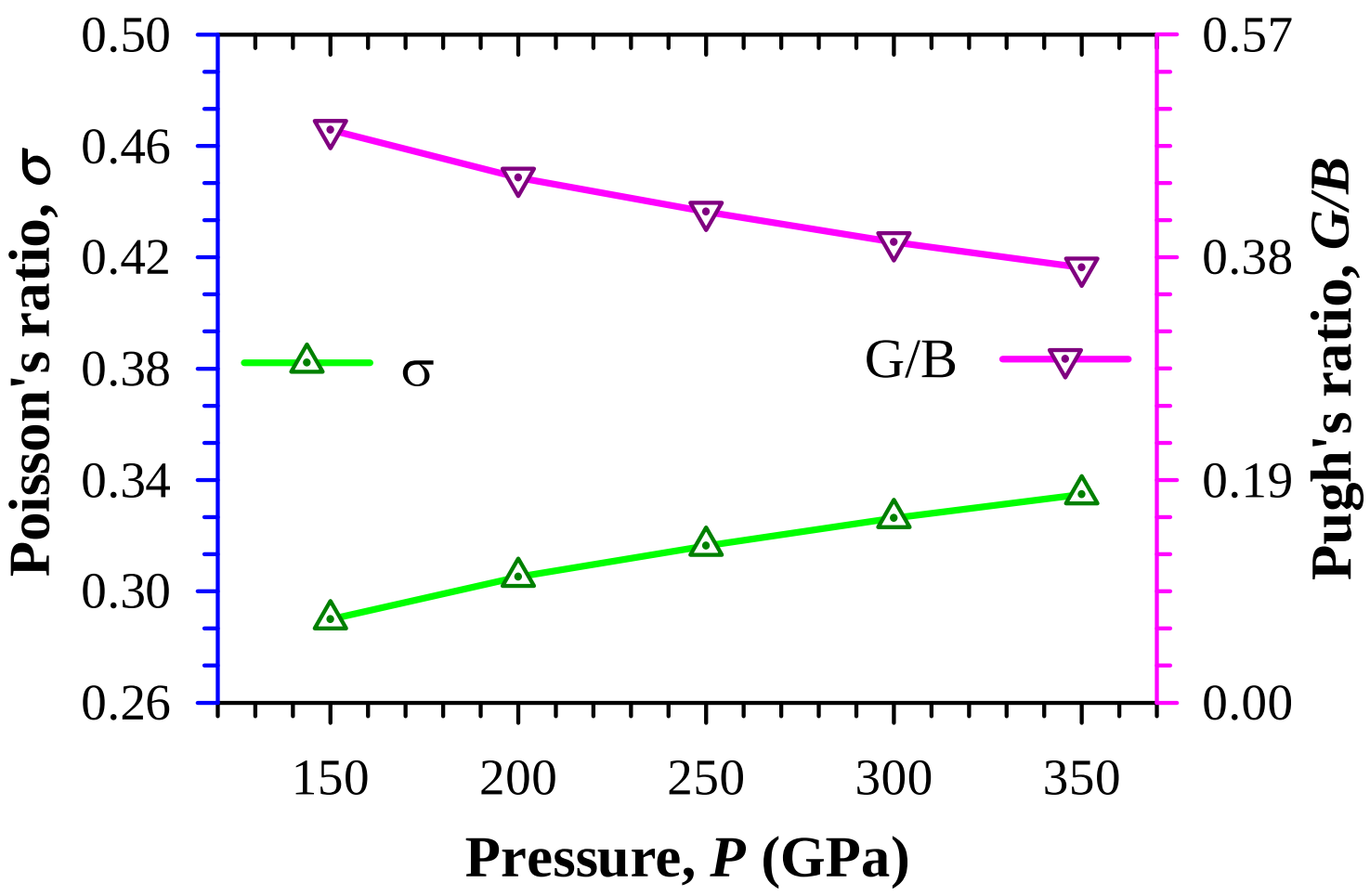

**Fig. 6**. Poisson's ratio ($\sigma$) and Pugh's ratio ($G/B$) as a function of pressure of $Li_5N$.

The Cauchy pressure, $P_C$ ($C_{13}$–$C_{44}$ and $C_{12}$–$C_{66}$ for hexagonal symmetry) can also be utilized to predict the bonding nature of solids. A positive $P_C$ designates the metallic bonding of materials with quasi-ductile nature, whereas the negative value endorses the strong directional covalent bonding with brittle nature [72,78]. As seen in **Fig. 7a**, the calculated Cauchy pressure are all positive within the studied range of pressure, indicating the metallic nature of $Li_5N$. However, increasing pressure from 150 GPa enhances ductility, which in turn strengthens plasticity in $Li_5N$ electride. Machinability index ($\mu_M$) defines a substance's cutting power, the most efficient methods for machine usage, and its plastic strain characteristics. $\mu_M$ escalates from 2.16 (150 GPa) to 2.80 (350 GPa) (**Table 3** and **Fig. 7b**). The highest $\mu_M$ found at 350 GPa indicates that the material is considerably more lubricating, has less friction, and maximizes production efficiency among all pressures, significantly affecting the production process [83].

Kleinman parameter $\xi$ is one of the mechanical tools to forecast the relative positions of the cation and anion sublattices under volume-conserving strain distortions. It can be calculated as: $\xi = (C_{11} + 8C_{12})/(7C_{11} + 2C_{12})$ [84]. Typically, a low value of $\xi$ indicates strong resistance to bond bending or bond-angle distortion, and vice-versa, *i.e.*, the bond bending (bond stretching) will be minimum when $\xi = 0$ ($\xi = 1$) [85]. The calculated Kleinman parameter of $Li_5N$ as a function of pressure is depicted in **Fig. 7b**. The increase of $\xi$ with rising pressure suggests that bond stretching is influenced more by lower pressures, while bending is affected by higher pressures.

Lamé moduli (first and second, represented by $\lambda$ and $\mu$, respectively) are important properties of materials that helps to understand how they respond to stress and strain. Generally, they are referred to as first ($\lambda$, measures the compressibility) and second Lamé's coefficient ($\mu$, quantifies the shear stiffness of material), respectively [86]. These are calculated as: $\lambda = E\sigma/(1 + \sigma)(1 - 2\sigma)$ and $\mu = E/2(1 + \sigma)$. It is worth noting that both coefficients are directly proportional to Young's modulus in which the latter one is nothing more than the shear modulus $G$ (cf. **Table 3**); therefore, $\mu = G$. Meanwhile, the simplified conditions for isotropic crystal are: $\lambda = C_{12}$ and $\mu = (C_{11}\text{-}C_{12})/2$ [87]. The material studied can be considered anisotropic, as the second coefficient does not meet the aforementioned criteria. As shown in **Fig. 7c**, both $\lambda$ and $\mu$ increase rapidly with rising pressure, indicating enhanced toughness of $Li_5N$ under higher pressure.

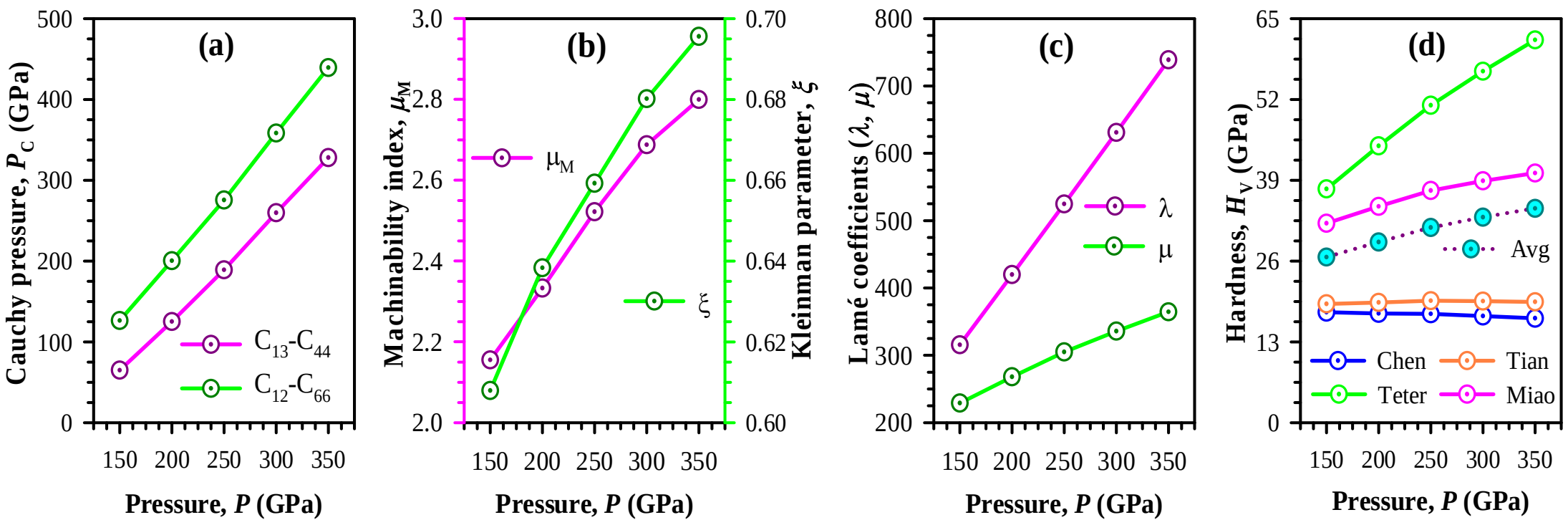

**Fig. 7**. (a) Cauchy pressure ($P_C$), (b) machinability index ($\mu_M$) and Kleinman parameter ($\xi$), (c) Lamé coefficients ($\lambda$ and $\mu$), and (d) Vickers harnesses ($H_V$) as a function of pressure for $Li_5N$ electride.

The polycrystalline bulk and shear moduli are still the essential parameters for assessing hardness. When a material demonstrates high hardness, both of these moduli tend to be significantly large, confirming its robust nature [88]. Mechanical hardness of crystals is often measured experimentally by indentation and is controlled either intrinsically or extrinsically. As a result, a universal and rigorous correlation between these conceptually distinct aspects of mechanical behavior cannot be expected [89]. Nonetheless, a number of semi-empirical correlations between elastic moduli and hardness have been postulated throughout the past few decades. Among them, four most-popular semi-empirical formula are used in this work to compute Vickers hardness ($H_V$) as follows [90–93]:

$$\left.\begin{aligned} H_V^{\text{Chen}} &= 2\left(\frac{G^3}{B^2}\right)^{0.585} - 3; \\ H_V^{\text{Tian}} &= 0.92\left(\frac{G}{B}\right)^{1.137} G^{0.708}; \\ H_V^{\text{Miao}} &= \frac{(1-2\sigma)Y}{6(1+\sigma)}; \\ H_V^{\text{Teter}} &= 0.1769G - 2.899. \end{aligned}\right\} \tag{8}$$

The computed average Vickers hardness values of $Li_5N$ increase from 26.62 GPa at 150 GPa pressure to 31.37 GPa and 34.48 GPa at 250 and 350 GPa, respectively (cf. **Table 3** and **Fig. 7d**). Since Vickers hardness is closely related to the shear resistance of a crystal, the increasing hardness suggests a corresponding strengthening of the directional bonding within the structure. A material is categorized as hard when $H_V$ exceeds 10 GPa [94] and superhard when $H_V$ exceeds 40 GPa [88]. The hardness value of 34.48 GPa at 350 GPa approaches the commonly accepted threshold of superhard materials demonstrating that $Li_5N$ exhibits remarkable mechanical robustness under compression. The combination of high hardness and pressure-induced strengthening makes it a promising candidate for applications requiring wear resistance and mechanical stability in high-pressure environments.

### *3.3. Elastic anisotropy*

The anisotropy of solids plays a crucial role in assessing mechanical durability and predicting micro-hardness. A significant elastic anisotropy may cause the induction of microcracks in the materials [95]. The elastic anisotropy for hexagonal crystal can be quantified by three shear anisotropic factors: $A_1$, $A_2$ and $A_3$. For {100}, {010}, and {001} shear planes between ⟨011⟩ and ⟨010⟩ directions, ⟨101⟩ and ⟨001⟩ directions, and ⟨110⟩ and ⟨010⟩ directions, respectively, these indices are evaluated from [72]:

$$\left.\begin{aligned} A_1 &= \frac{C_{11} + C_{12} + 2C_{33} - 4C_{13}}{6C_{44}};\ A_2 = \frac{2C_{44}}{C_{11} - C_{12}}; \\ A_3 &= A_1 A_2 = \frac{C_{11} + C_{12} + 2C_{33} - 4C_{13}}{3(C_{11} - C_{12})}. \end{aligned}\right\} \tag{9}$$

Percent anisotropy in compression ($A_B$) and shear ($A_G$) introduced by Chung and Buessem, along with the generalized Zener anisotropy index often known as universal anisotropic index ($A^U$) are also vital anisotropy indices [96]. They are evaluated as [72]:

$$A_B = \frac{B_V - B_R}{B_V + B_R} \times 100\%;\ A_G = \frac{G_V - G_R}{G_V + G_R} \times 100\%;\ A^U = \frac{5G_V}{G_R} + \frac{B_V}{B_R} - 6. \tag{10}$$

For an isotropic system, $A_1$, $A_2$ and $A_3$ should be unity, and any deviation from unity measures the degree of anisotropy. In contrast, $A_B = A_G = 0$ for an isotropic structure, and the index $A^U$ has either zero or positive value; a zero value represents a perfectly isotropic nature, whereas a positive value signifies a specific level of anisotropy in elasticity of solids. Moreover, the ratio of linear compressibility coefficients along $c$-to-$a$ axis for hexagonal crystals is given by, $f = (k_c/k_a) = (C_{11} + C_{12} - 2C_{13})/(C_{33} - C_{13})$, where $k_a$ and $k_c$ are the linear compressibility coefficients along the $a$- and $c$-axis, respectively. Isotropic compressibility is achieved when $f = 1$, while the deviation from the unity measures the degree of anisotropy [72].

**Table 4**. Calculated shear anisotropic factors $A_1$, $A_2$, and $A_3$, universal anisotropy index $A^U$ and percentage anisotropy factors $A_B$ (in %) and $A_G$ (in %), and ratio of linear compressibility coefficient $f$ of $Li_5N$ at various pressures.

| $P$ (GPa) | $A_1$ | $A_2$ | $A_3$ | $A_B$ | $A_G$ | $A^U$ | $f$ |
|---|---|---|---|---|---|---|---|
| 150 | 1.260 | 0.987 | 1.243 | 0.015 | 0.407 | 0.041 | 1.061 |
| 200 | 1.246 | 1.001 | 1.247 | 0.012 | 0.389 | 0.039 | 1.057 |
| 250 | 1.273 | 0.985 | 1.254 | 0.017 | 0.442 | 0.045 | 1.069 |
| 300 | 1.283 | 0.989 | 1.268 | 0.012 | 0.477 | 0.048 | 1.060 |
| 350 | 1.272 | 1.004 | 1.270 | 0.012 | 0.516 | 0.052 | 1.058 |

It is indicative from **Table 4** and **Fig. 8** that all indices demonstrate the anisotropic signature of $Li_5N$ within the studied pressure range. Among these, the values of $A_2$ are very close to 1 throughout the pressure range studied, indicating quasi-isotropic nature in $Li_5N$ along {010} shear plane. As a whole, all the estimated indices exhibit almost an invariant effect with fractional anomaly while applying pressure from 150 GPa to 350 GPa.

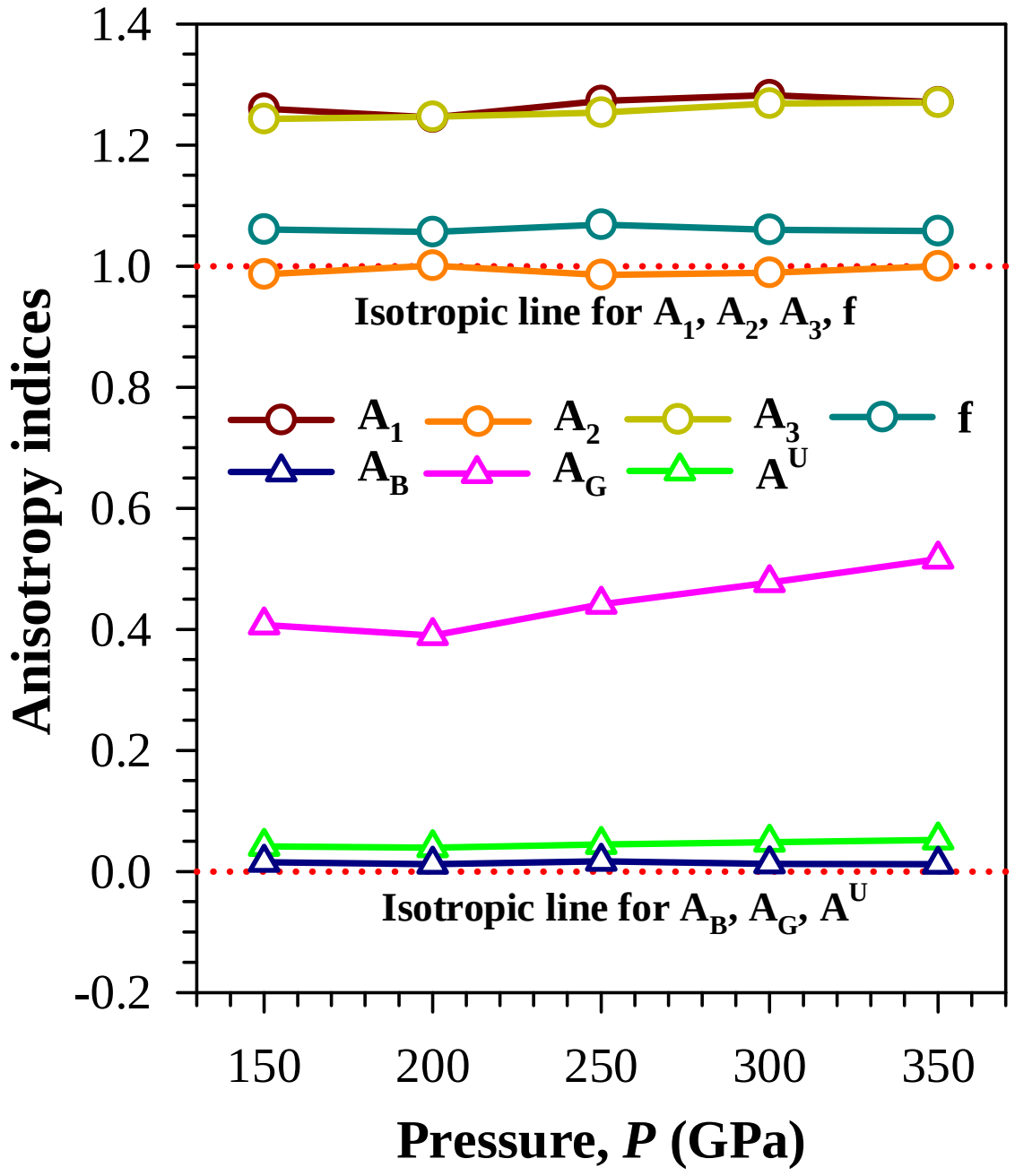

**Fig. 8**. (a) Shear ($A_1$, $A_2$, and $A_3$), universal ($A^U$) and percentage ($A_B$ and $A_G$ (in %) anisotropy factors, and linear compressibility coefficients ratio (*f*) as a function of pressure for $Li_5N$.

Since the elastic anisotropy is directional, the elastic anisotropies need to be described using the three-dimensional (3D) surface construction. To analyze the anisotropy of $Li_5N$, we used 3D plots in view of elastic moduli *i.e.*, Young's modulus (*E*), linear compressibility (*β*), shear modulus (*G*), and Poisson's ratio (*ν*). This analysis was conducted using VELAS software [97] at pressures of 150 GPa, 250 GPa, and 350 GPa. A spherical shape designates perfect isotropy and any deviation from the sphere exhibits the degree of anisotropy. **Table 5** summarizes the maximum and minimum values of the moduli and their corresponding anisotropy level and **Fig. 9** depicts the 3D plots of elastic moduli; and these plots clearly indicate their anisotropy. The inclusion of pressure enhances the level of anisotropy in $A_E$ and $A_G$ along with an anomaly in case of $A_\beta$ and $A_\nu$. Overall, the minimum anisotropy is achieved in $Li_5N$ by means of linear compressibility across all indices.

**Table 5**. Minimum and maximum values of elastic moduli (*E* in GPa, *β* in $TPa^{-1}$, *G* in GPa, *ν*) and their corresponding elastic anisotropy ($A_E$, $A_\beta$, $A_G$, $A_\nu$) for $Li_5N$ at various pressures.

| *P* (GPa) | $E_{min}$ | $E_{max}$ | $A_E$ | $\beta_{min}$ | $\beta_{max}$ | $A_\beta$ | $G_{min}$ | $G_{max}$ | $A_G$ | $\nu_{min}$ | $\nu_{max}$ | $A_\nu$ |
|---|---|---|---|---|---|---|---|---|---|---|---|---|
| 150 | 566.49 | 678.41 | 1.198 | 0.698 | 0.740 | 1.061 | 217.30 | 258.04 | 1.187 | 0.220 | 0.361 | 1.643 |
| 250 | 768.00 | 931.22 | 1.213 | 0.448 | 0.478 | 1.069 | 288.71 | 345.43 | 1.196 | 0.243 | 0.392 | 1.613 |
| 350 | 927.33 | 1149.90 | 1.240 | 0.333 | 0.353 | 1.061 | 343.65 | 417.13 | 1.214 | 0.255 | 0.416 | 1.627 |

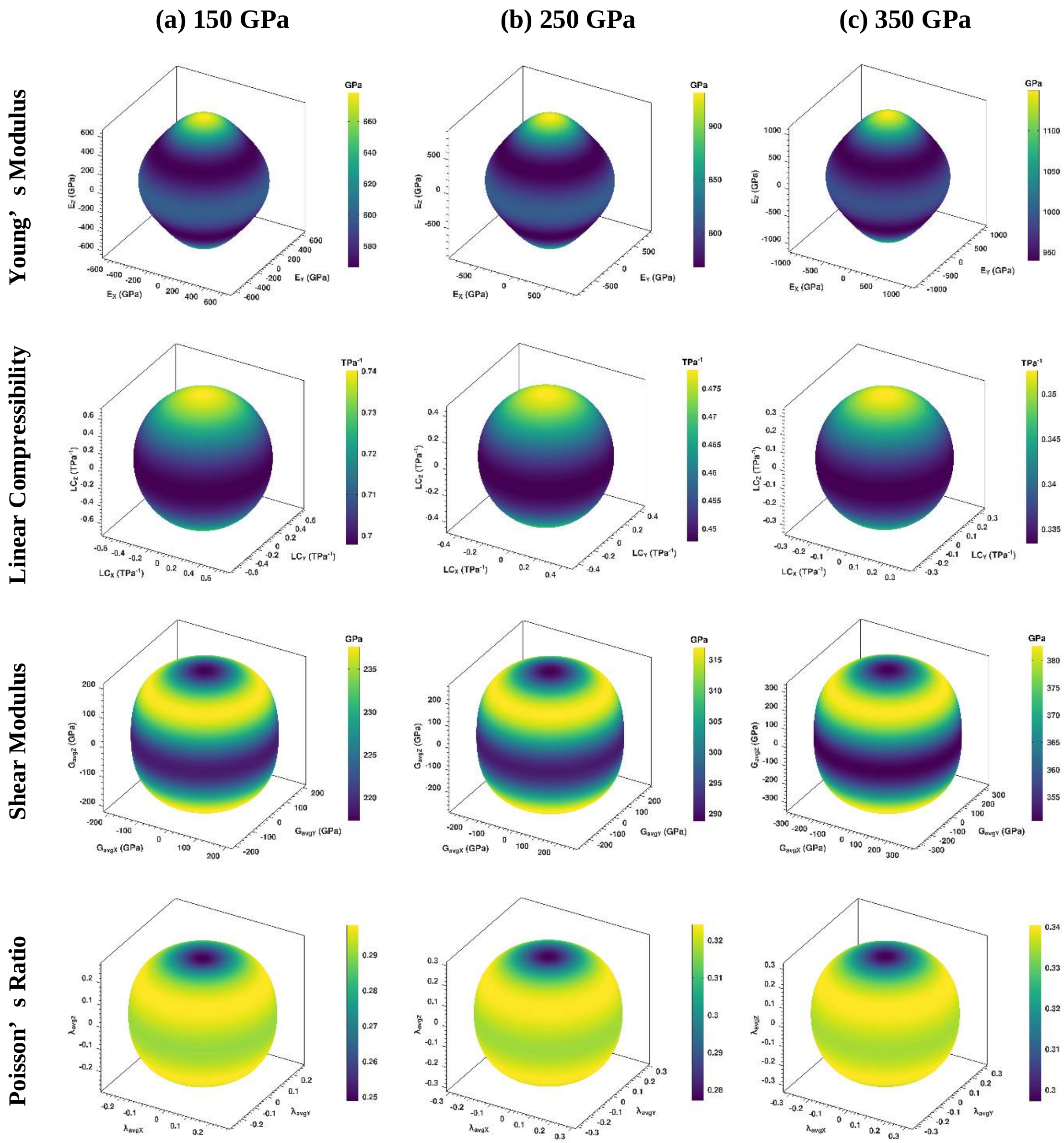

**Fig. 9**. Direction-dependent contour plots of Young’s modulus (*E*), linear compressibility (*β*), shear modulus (*G*) and Poisson’s ratio (*σ*) of $Li_5N$ at 150 GPa, 250 GPa and 350 GPa.

### *3.4. Debye temperatures and anisotropy in sound velocities*

A fundamental physical parameter, known as the Debye temperature ($\Theta_D$), is closely related to many physical properties, including specific heat and melting temperatures. It can be evaluated using the Anderson model [98] as follows:

$$\Theta_D = \frac{h}{k_B}\left[\frac{3n}{4\pi}\left(\frac{N_A\rho}{M}\right)\right]^{\frac{1}{3}} v_m;\ v_m = \left[\frac{1}{3}\left(\frac{2}{v_t^3} + \frac{1}{v_l^3}\right)\right]^{-\frac{1}{3}};\ v_l = \left(\frac{3B + 4G}{3\rho}\right)^{\frac{1}{2}};\ v_t = \left(\frac{G}{\rho}\right)^{\frac{1}{2}}, \quad (11)$$

where $h$ and $k_B$ are the Planck's and Boltzmann's constants, respectively, $N_A$ is the Avogadro's number, $\rho$ is the density of the crystal, $M$ is the molecular weight, and $n$ is the number of atoms contained in one molecular formula. In **Eqn. 11**, $v_m$ denotes the average sound velocity that can obtained from longitudinal ($v_l$) and transverse velocity ($v_t$), as estimated by Navier's equation [99].

Acoustic velocities serve as important indicators of the thermal properties of solids, providing valuable insights for further exploration and understanding of material behavior. According to **Table 6**, the calculated sound velocities and Debye temperatures increase with pressure. This suggests that higher pressure leads to hardening of $Li_5N$, as supported by the Vickers hardness values presented in **Section 3.2**. The longitudinal velocities are always higher than transverse/shear one in each pressure. Moreover, high values of Debye temperatures (ranging from ~1764 K to ~2134 K) indicate that $Li_5N$ is expected to be highly thermally conductive at high pressures. Furthermore, the highest bond strength in $Li_5N$ is anticipated to occur at 350 GPa, which corresponds to the highest $\Theta_D$; conversely, the weakest bond strength is expected at 150 GPa [79].

**Table 6**. Calculated longitudinal, transverse and average ($v_l$, $v_t$, and $v_m$ in km/s) sound velocities, Debye temperatures ($\Theta_D$ in K), and anisotropic sound velocities (in kms$^{-1}$) for $Li_5N$ electride at various pressures.

| $P$ | $v_l$ | $v_t$ | $v_m$ | $\Theta_D$ | <100> | | | | <001> | | | |
|---|---|---|---|---|---|---|---|---|---|---|---|---|
| | | | | | $v_l$ | $v_{t_1}$ | $v_{t_2}$ | $v_m$ | $v_l$ | $v_{t_1}$ | $v_{t_2}$ | $v_m$ |
| 150 | 16.162 | 8.791 | 9.807 | 1763.93 | 16.304 | 3.298 | 8.567 | 6.737 | 16.630 | 8.567 | 8.567 | 9.593 |
| 250 | 17.975 | 9.318 | 10.429 | 1985.53 | 18.134 | 3.596 | 9.068 | 7.197 | 18.456 | 9.068 | 9.068 | 10.182 |
| 350 | 19.270 | 9.603 | 10.775 | 2133.92 | 19.427 | 3.809 | 9.419 | 7.522 | 19.821 | 9.324 | 9.324 | 10.596 |

The anisotropy in sound velocities can be characterized by estimating direction-dependent sound velocities as a function of pressure within the Christoffel's equation [100]. The solution of this equation comprises two parts; *e.g*., one longitudinal ($v_l$) and two transverse modes - first transverse ($v_{t_1}$) and second transverse ($v_{t_2}$). There are <100> and <001> directional sound velocities in hexagonal symmetry which are estimated as follows [101,102]:

$$\left.\begin{array}{l} v_l^{<100>} = \left(\frac{C_{11}}{\rho}\right)^{\frac{1}{2}}, v_{t_1}^{<100>} = \left(\frac{C_{11}-C_{12}}{2\rho}\right)^{\frac{1}{2}}, v_{t_2}^{<001>} = \left(\frac{C_{44}}{\rho}\right)^{\frac{1}{2}}; \\ v_l^{<001>} = (\frac{C_{33}}{\rho})^{\frac{1}{2}}, v_{t_1}^{<001>} = v_{t_2}^{<001>} = (\frac{C_{44}}{\rho})^{\frac{1}{2}}. \end{array}\right\} \quad (12)$$

From **Table 6**, it is clear that longitudinal acoustic velocities are always higher than transverse/shear in all directions at an identical pressure; among these, the maximum velocities are obtained at 350 GPa. Also, the largest $v_l$, $v_{t_1}$, $v_{t_2}$are exhibited at 350 GPa may be due to the relatively large $C_{33}$ and lowest $c/a$ ratio at this pressure. The data presented in **Table 6** clearly demonstrates that $v_{t_1}$ and $v_{t_2}$ are distinctly different along the <100> direction. Additionally, $v_l$ in the <001> direction surpasses $v_l$ in the <100> direction at a specific pressure. This evidence decisively confirms that the varying sound velocities in different directions contribute to the pronounced anisotropic nature of $Li_5N$ under high pressure.

*3.5. Thermal conductivities*

*3.5.1. Minimum thermal conductivities and anisotropy*

The thermal conductivity ($k$) of a material reflects its capacity to conduct heat. As the temperature increases, there comes a point where the thermal conductivity reaches a minimum value, known as minimum thermal conductivity ($k_{\mathrm{m}}$). Understanding this concept can help in selecting materials for applications where efficient heat transfer is critical. Cahill and Pohl (*C-P*) proposed an empirical model to evaluate $k_{\mathrm{m}}$ of a crystal at low temperatures that comprises both isotropic and anisotropic cases as follows [103]:

$$k_m^{C-P} = \begin{cases} \frac{1}{2.48} k_B n_v^{\frac{2}{3}} (v_l + 2v_t) & \text{(Isotrpoic);} \\ \frac{1}{2.48} k_B n_v^{\frac{2}{3}} (v_l + v_{t_1} + v_{t_2}) & \text{(Anisotrpoic),} \end{cases} \quad (13)$$

where $n_v$ denotes the number of atoms per unit volume and $k_{\mathrm{B}}$ is the Boltzmann's constant. For isotropic solids, the two shear wave velocities are identical. Later on, Clarke (*C*) proposed another model to evaluate $k_{\mathrm{m}}$ for isotropic solids as [104]:

$$k_m^C = k_B v_m \left(\frac{M}{n\rho N_A}\right)^{-\frac{2}{3}}. \quad (14)$$

**Table 7**. Calculated isotropic and anisotropic minimum thermal conductivities ($k_m$ in $\mathrm{Wm^{-1}K^{-1}}$), acoustic Grüneisen constant ($\gamma_{\mathrm{a}}$), lattice thermal conductivities at room temperature ($k_l$ in $\mathrm{Wm^{-1}K^{-1}}$), and melting temperatures ($T_{\mathrm{m}}$ in K) for $Li_5N$ electride at high pressures.

| $P$ | $k_m^{C-P}$ | $k_m^C$ | $k_m^{C-P}$ along | | $\gamma_{\mathrm{a}}$ | $k_l$ | $T_{\mathrm{m}}$ |
|---|---|---|---|---|---|---|---|
| | | | <100> | <001> | | | |
| 150 | 15.255 | 10.994 | 12.734 | 15.263 | 1.712 | 0.986 | 5156.27 |
| 250 | 18.542 | 13.099 | 15.598 | 18.533 | 1.878 | 1.792 | 7391.01 |
| 350 | 21.088 | 14.644 | 17.897 | 21.084 | 2.011 | 2.513 | 9456.70 |

The estimated minimum thermal conductivities ($k_{\mathrm{m}}$) for both isotropic and anisotropic conditions for $Li_5N$ are listed in **Table 7**. As seen, $k_{\mathrm{m}}$ are always found higher in *Cahill-Pohl* method than that of *Clarke* method. This might stem from the fact that the optical phonon spectrum is overlooked in the *Clarke* model, while it is included in the *Cahill-Pohl* model [61]. It is seen that $k_{\mathrm{m}}$ values escalate along 150 GPa→350 GPa in both models. In addition, the estimated $k_{\mathrm{m}}$ along <100> directions are clearly dissimilar from that along <001> at identical pressure signifies the anisotropic nature of $Li_5N$.

*3.5.2. Lattice thermal conductivities and melting temperatures*

The lattice thermal conductivity ($k_l$) quantifies how much heat energy is transported by the lattice vibration when there is a temperature gradient in the system. It serves as an important indicator of the anharmonic characteristics of a structure, allowing us to effectively assess the potential of materials for thermoelectric applications [68]. It can be estimated as a function of temperature within the well-known Slack's model [105] as:

$$k_l = A\frac{M_{av}\Theta_D^3\delta}{{\gamma_a}^2 n^{\frac{2}{3}}T}, \tag{15}$$

where $M_{av}$ is the average atomic mass in kg/mol, $\Theta_D$ is the Debye temperature in K, $\delta$ is the cubic root of the average atomic volume in meter, and $\gamma_a$ refers to the acoustic Grüneisen constant: $\gamma_a = 1.5(1+\sigma)/(2-3\sigma)$, $n$ is the number of atoms in the unit cell, and $T$ is the absolute temperature in K, and the factor $A$ ($\gamma_a$) is a constant which can be calculated according to Julian [106] as: $A(\gamma_a) = 4.85628 \times 10^7 / 2(1 - 0.514{\gamma_a}^{-1} + 0.228{\gamma_a}^{-2})$.

Understanding the melting temperature ($T_m$) of solids is crucial for the development of effective high-temperature devices. This knowledge can greatly enhance the design and performance of materials used in such applications. $T_m$ for hexagonal solid is estimated roughly as [107]: $T_m = 354 + 1.5(2C_{11} + C_{33})$.

The calculated lattice thermal conductivities ($k_l$) at room-temperature ($T$ = 300 K) together with acoustic Grüneisen parameter, and melting temperatures are tabulated in **Table 7**. As seen, all three indices are intensified with pressures in accordance to sound velocities and Debye temperature. At room-temperature, $k_l$ varies from 0.986 $Wm^{-1}K^{-1}$ (at 150 GPa) to 2.513 $Wm^{-1}K^{-1}$ (at 350 GPa). It is seen that pressure has significant effect on lattice thermal conductivity of $Li_5N$.

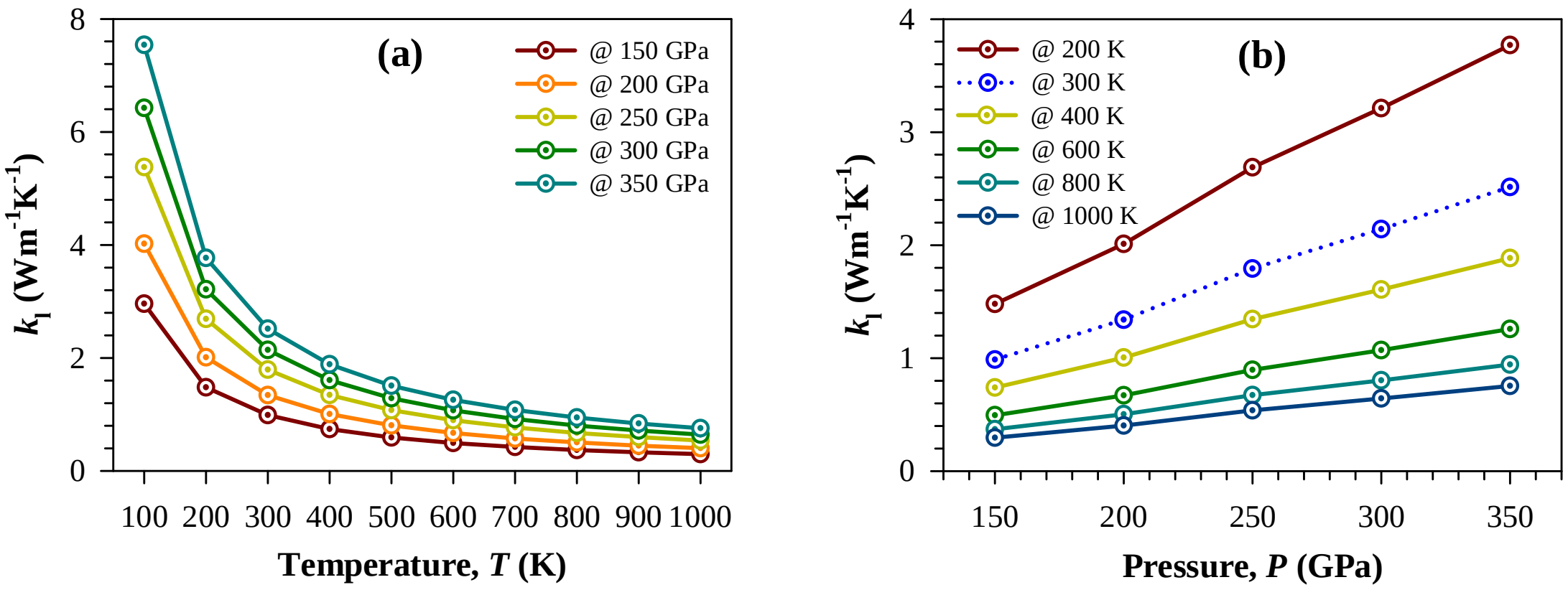

**Fig. 10**. Lattice thermal conductivity of $Li_5N$ as a function of (a) temperature and (b) pressure.

Since $Li_5N$ exhibits high-temperature phase stability, as studied by Z. Wan *et al.* via *ab initio* molecular dynamics (AIMD) simulations [26], it is necessary to calculate $k_l$ at high temperatures at various pressures. The temperature (from 100 K to 1000 K) and pressure dependence on the lattice thermal conductivities ($k_l$) at various pressures and temperatures is shown in **Fig. 10a** and **Fig. 10b**, respectively. According to **Fig. 10a**, a gradual decrease in lattice thermal conductivity is observed as temperature increases at each pressure, whereas **Fig. 10b** portrays a monotonically increasing trend of $k_l$ with pressure at a particular temperature.

It is important to note that at higher temperatures, $k_l$ tends to approach a saturation point. We have not found any experimental or theoretical data regarding lattice thermal conductivities for $Li_5N$; therefore, this work may serve as a useful guideline for upcoming research.

The acoustic Grüneisen constant ($\gamma_a$) measures the anharmonic behavior of interatomic interactions of solids. Among various types, Grüneisen constants arising from acoustic ($\gamma_a$), elastic ($\gamma_e$), and thermodynamic ($\gamma_d$) considerations are identical for metallic, ionic, and molecular systems [73]. The higher $\gamma_a$ designates the higher anharmonicity and lower phonon thermal conductivity. The calculated values of $\gamma_a$ varies from 1.712 to 2.011 from 150 GPa to 350 GPa in $Li_5N$; *i.e.*, the increment of pressure enhances the anharmonicity. The melting temperature $T_m$ increases from 5156.27 to 9456.70 K with pressure from 150 GPa to 350 GPa, indicating that $Li_5N$ possesses a stiffer lattice and good thermal conductivity along 150 GPa→350 GPa. The melting temperatures are extremely high within the pressure range considered.

### *3.6. Electronic properties*

#### *3.6.1. Band structure and density of states*

Band structure reveals electronic energy dispersion properties. The electronic nature of $Li_5N$ under 150-350 GPa pressure are calculated within the bulk electronic band structure, total density of states (TDOS) and partial density of states (PDOS). Band structure calculations were also performed using Quantum Espresso compiled with Winmoster software, taking into account the spin-orbit coupling (SOC) effect. However, our analysis showed that the SOC has a minimal impact on the results compared to those obtained without SOC, as the lighter elements Li and N are responsible in forming $Li_5N$. Since there was no notable difference between the two sets of results, we decided to focus only on the case without the SOC effect in this study. Moreover, the spin polarization was deliberately excluded due to the nonmagnetic nature of the compound, and DFT + U was not necessary, as these materials do not possess highly localized *d* or *f* electrons. The electronic energy dispersion curves of $Li_5N$ along high symmetry directions (*Γ-A-H-K- Γ-M-L-H*) of the first BZ in the energy range from −20 eV to +20 eV at 150 GPa, 250 GPa, and 350 GPa pressures are depicted in **Fig. 11a–c**.

The electronic band structure of $Li_5N$ reveals the absence of an energy band gap at the Fermi level ($E_F$), indicating intrinsic metallic conductivity. In metallic systems, electrons can be excited to nearby unoccupied states with negligible energy cost, eliminating the need for thermal activation. The metallic behavior primarily arises from significant Li-*s* and N-*p* orbital hybridization around the Fermi level, $E_F$. This interaction generates highly dispersive bands that extend across the Fermi level and overlap continuously, preventing gap formation. The resulting finite density of states at $E_F$ provides a sufficient population of mobile charge carriers, thereby facilitating efficient electronic transport throughout the crystal. Such orbital hybridization and band overlap are characteristic features of metallic compounds and play a crucial role in determining their electrical, thermal, and superconducting properties. The increase in pressure beyond 150 GPa does not induce opening a band gap, indicating that the compound retains its metallic nature throughout the investigated pressure range although it significantly alters the dispersion and density of electronic states around the Fermi level, thereby impacting the degree of metallicity. These pressure-induced changes in the electronic structure influence the density of states (DOS) at $E_F$ and the availability of charge carriers, which can have important consequences for the transport, bonding, and superconducting properties of the material.

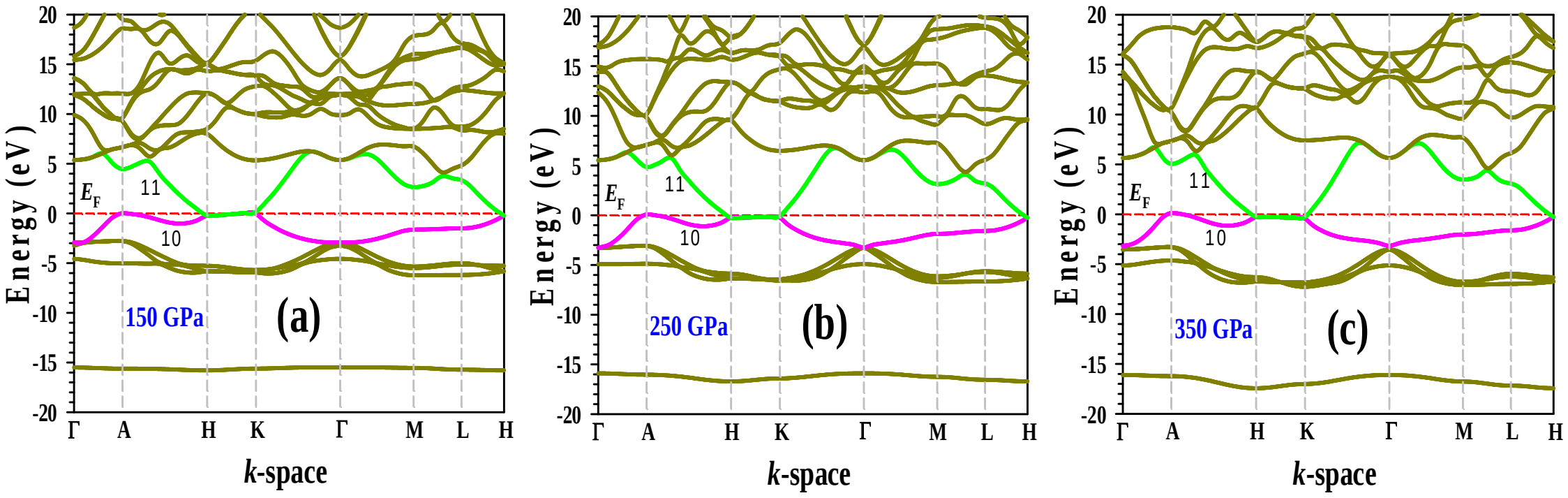

**Fig. 11**. Electronic band structures of $Li_5N$ at (a) 150 GPa, (b) 250 GPa and (c) 350 GPa pressures.

The metallic behavior of $Li_5N$ electride is further corroborated by the TDOS and PDOS, which exhibits small but finite values at the Fermi level for all pressures. The top panel of **Fig. 12a** shows the TDOS for all compositions whereas the other panels represent the orbital-resolved PDOS for individual atom.

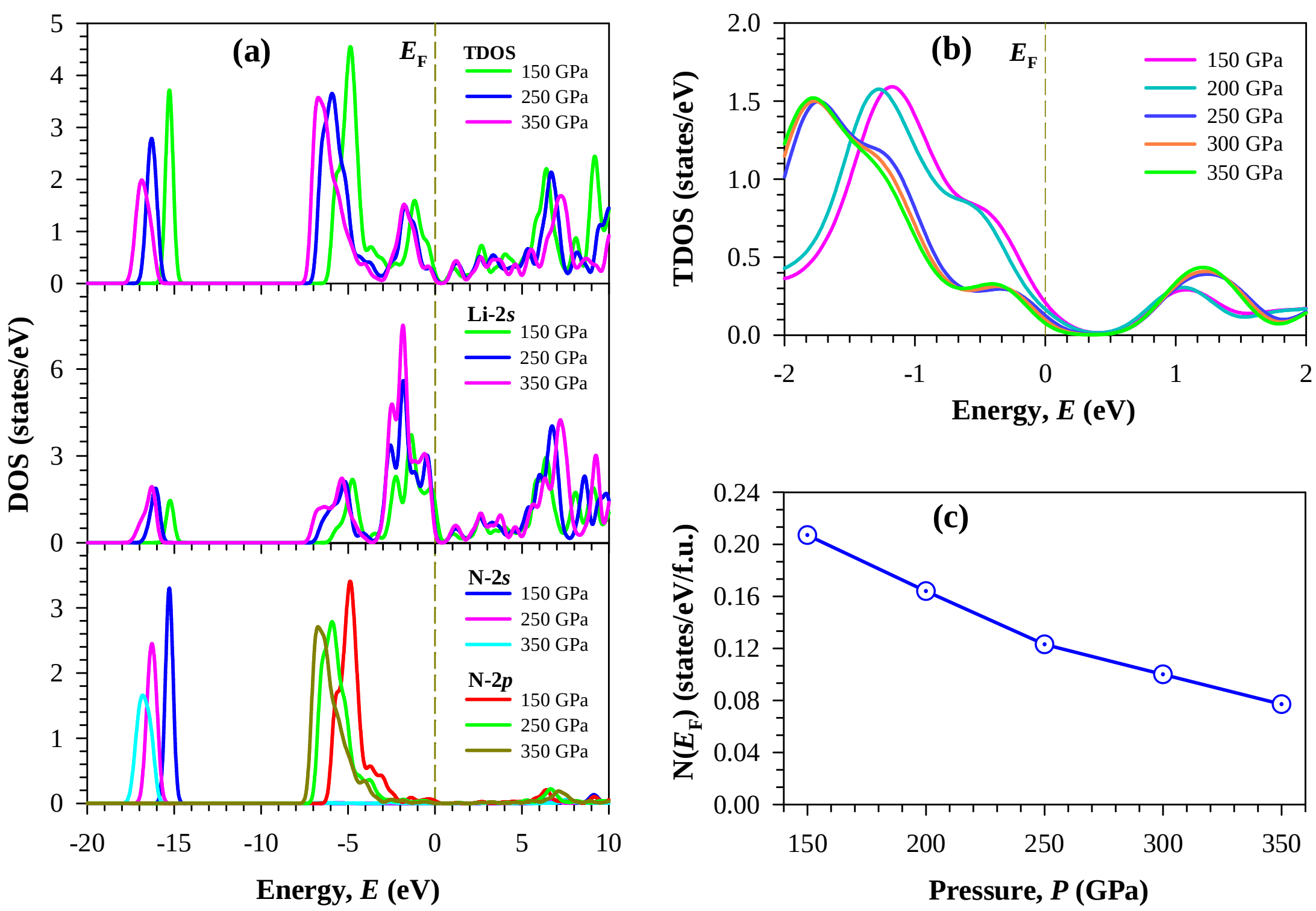

**Fig. 12.** Calculated (a) total (T) and partial (P) density of states (DOS) at various pressures, (b) projected TDOS near $E_F$ and (b) TDOS at $E_F$ as a function of pressure for $Li_5N$.

In terms of partial density of states (PDOS), the DOS in the valence band (about -20 to -12 eV) is primarily governed by Li-*s* with significant contributions from N-*s* states and minimal contributions from N-*p* states. The substantial overlap between Li-*s* and N-*s* orbitals in this energy range signifies considerable hybridization. In the region from -8 eV to -4 eV, DOS is attributed due to the N-*p* states in association with Li-*s* states. Near the $E_F$, from -3 eV to 0 eV, DOS is strongly influenced by Li-*s* states. Therefore. the TDOS at $E_F$ reaches a maximum of 0.207 states/eV at 150 GPa, then diminishes to a minimum of 0.077 states/eV at 350 GPa (**Fig. 12b**). The systematic decline in DOS at $E_F$ from 150 GPa to 350 GPa (**Fig. 12c**) reflects the increase in orbital dispersion due to lattice contraction, consistent with the Drude-like intraband optical response and high low-energy reflectivity observed in the optical spectra (discussed later in **section 3.8**). Moreover, the progressive decrease in $N(E_F)$ with increasing pressure indicates a reduction in the number of electronic states available at $E_F$. This behavior is attributed to pressure-induced band broadening arising from enhanced orbital overlap under compression. Consequently, the metallic character becomes less pronounced, and the reduced $N(E_F)$ weakens the electron–phonon coupling strength, resulting to the observed suppression of superconductivity at higher pressures [108].

### *3.6.2. Mulliken population analysis*

Analysis of Mulliken's atomic population (MAP) is useful to explain bond overlap population (BOP), charge transfer mechanism and effective valence charge (EVC). Having zero/negative values of BOP signifies either negligible atomic interactions or emphasize anti-bonding states, and hence indicates a remarkably weak bonding that can be ineffective in estimating theoretical hardness [78]. A large (small) positive value of BOP indicates the high degree of covalency (iconicity). In contrast, EVC can be measured from the difference between the formal ionic charge and Mulliken charge within a solid which is used to predict the bonding character of materials. A perfect ionic bonding appears if the value of EVC is zero, and any deviation from zero intensifies the strength of covalent bonding [109]. The calculated Mulliken atomic populations along with effective valence charge of $Li_5N$ at various pressures is tabulated in **Table 8**. The spilling parameter quantifies the fraction of electronic states that cannot be represented by the localized atomic-orbital basis set used for population analysis. Lower spilling values indicate a more complete and accurate projection of the plane-wave wavefunctions onto the atomic basis. All bands spilling parameter for Mulliken analysis are 0.07%, 0.07%, 0.08% for 150 GPa, 250 GPa and 350 GPa, respectively. Such small values demonstrate an excellent projection quality and validate the reliability of the calculated Mulliken charges and bond populations over the entire pressure range [110].

**Table 8**. Calculated Mulliken atomic populations and effective valence charge (EVC) of $Li_5N$ at various pressures.

| *P* (GPa) | Species | Mulliken atomic population | | | Mulliken charge | Formal ionic charge | EVC |
|---|---|---|---|---|---|---|---|
| | | *s* | *p* | Total | | | |
| 150 | Li (1) | 1.92 | 2.5 | 4.42 | -1.42 | +1 | 2.42 |
| | Li (2) | 0.41 | 1.95 | 2.36 | 0.64 | +1 | 0.36 |
| | Li (3) | 0.41 | 1.95 | 2.36 | 0.64 | +1 | 0.36 |
| | Li (4) | 0.41 | 1.95 | 2.36 | 0.64 | +1 | 0.36 |
| | Li (5) | 0.41 | 1.95 | 2.36 | 0.64 | +1 | 0.36 |
| | N | 1.71 | 4.41 | 6.12 | -1.12 | -3 | -1.88 |
| 250 | Li (1) | 2.62 | 2.76 | 5.39 | -2.39 | +1 | 3.39 |
| | Li (2) | -0.06 | 2.16 | 2.1 | 0.90 | +1 | 0.10 |
| | Li (3) | -0.06 | 2.16 | 2.1 | 0.90 | +1 | 0.10 |
| | Li (4) | -0.06 | 2.16 | 2.1 | 0.90 | +1 | 0.10 |
| | Li (5) | -0.06 | 2.16 | 2.1 | 0.90 | +1 | 0.10 |
| | N | 1.78 | 4.43 | 6.21 | -1.21 | -3 | -1.79 |
| 350 | Li (1) | 3.06 | 2.92 | 5.98 | -2.98 | +1 | 3.98 |
| | Li (2) | -0.4 | 2.33 | 1.93 | 1.07 | +1 | -0.07 |
| | Li (3) | -0.4 | 2.33 | 1.93 | 1.07 | +1 | -0.07 |
| | Li (4) | -0.4 | 2.33 | 1.93 | 1.07 | +1 | -0.07 |
| | Li (5) | -0.4 | 2.33 | 1.93 | 1.07 | +1 | -0.07 |
| | N | 1.86 | 4.42 | 6.28 | -1.28 | -3 | -1.72 |

A positive Mulliken charge indicates that an atom transfers electrons away, while a negative charge means it accumulates electronic charge. It is clear from **Table 8** that the charge is transferred from Li to Li and Li to N. The charge transfer between Li and Li increases, while transfer between Li and N atoms decreases as the pressure increases. As Li and N atoms become farther apart, more charges are transferred between them at higher pressures, therefore, the ionic bonds in $Li_5N$ diminish at higher pressures, and subsequently covalency enhances. This conclusion is consistent with the analysis indicating an increasing Poisson's ratio under higher pressure [111].

### *3.6.3. Charge density and bonding nature*

The charge (/electron) density difference mapping (CDDM) is a valuable tool for elucidating the bonding characteristics and charge distribution within a material. **Figure 13** depicts the calculated CDD of $Li_5N$ at 150 GPa, 250 GPa, and 350 GPa. The color scale on the right denotes the CD difference values in e/Å$^3$, where red and blue regions correspond to electron accumulation and electron depletion, respectively. The CDDM provides insight into the redistribution of electronic charge upon bond formation, thereby revealing the nature and strength of the interatomic interactions in the compound.

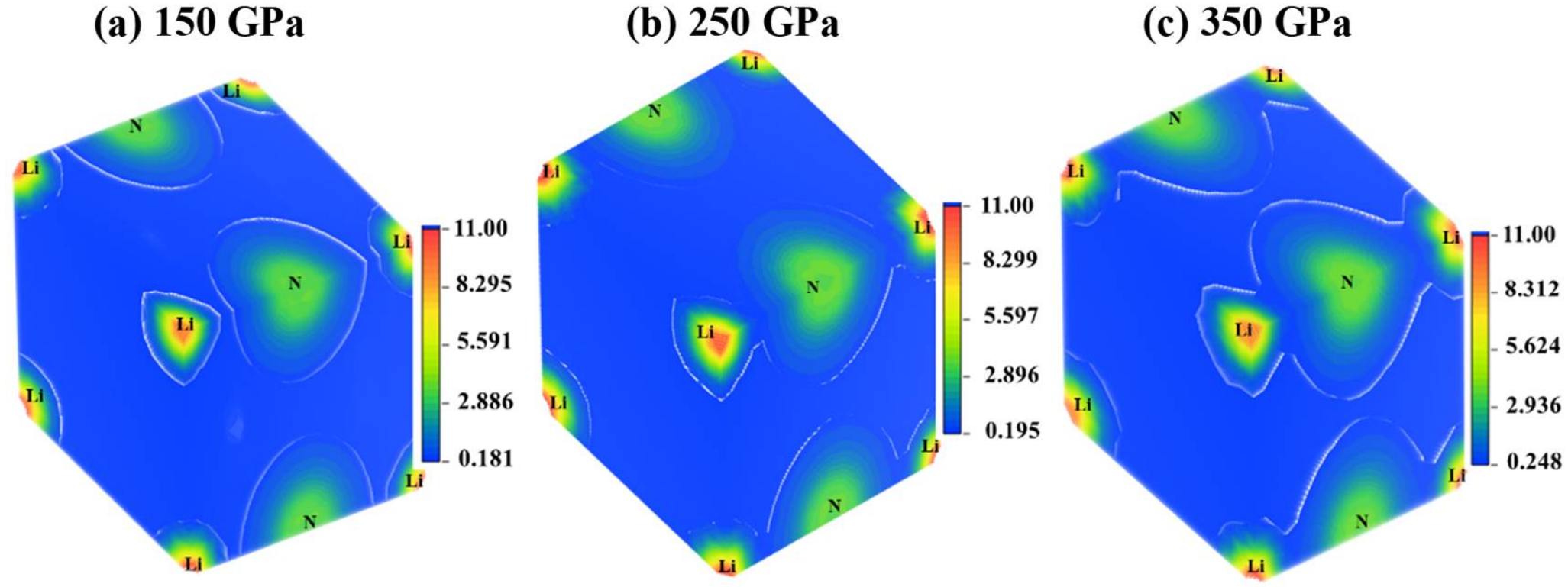

**Fig. 13**. Charge density mapping of $Li_5N$ for (a) 150 GPa, (b) 250 GPa, (c) 350 GPa.

It is evident from **Fig. 13** that charge is transferred from Li to N, as indicated by the electron accumulation around the N atoms and the corresponding electron depletion near the Li atoms. The noticeable overlap of electron density between Li and N suggests the presence of covalent bonding with partial ionic character. In contrast, the Li–Li interaction exhibits a predominantly ionic (or non-covalent) nature, as the electron densities associated with neighboring Li atoms do not significantly overlap or penetrate each other. This bonding characteristics are supported by the Mulliken atomic population (MAP) analysis, which indicates similar charge-transfer behavior and bonding scenarios.

*3.7. Superconducting properties*

The superconducting behavior of a material is fundamentally governed by its electronic structure, electron–phonon interactions, and many-body electronic correlations. Important parameters, including the electron–phonon coupling constant ($\lambda_{ep}$), the Coulomb pseudopotential ($\mu^*$), characteristic phonon energy, and the density of states at the Fermi level [$N(E_F)$], play crucial roles in determining the emergence and magnitude of superconductivity. These quantities directly affect the formation of Cooper pairs and strongly influence the superconducting transition temperature ($T_c$) as well as the overall strength of the superconducting state including the thermodynamic critical field and the superconducting energy gap.

The effective Coulomb interaction is described by the dimensionless Coulomb pseudopotential $\mu^*$, and can be calculated from the TDOS at $E_F$, using the phenomenological Bennemann-Garland formula [112]:

$$\mu^* = \frac{0.26\, N(E_F)}{1 + N(E_F)} \tag{16}$$

The electron–phonon coupling constant ($\lambda_{ep}$) can be calculated using the familiar McMillan equation [113] when $T_c$, Debye temperature and $\mu^*$ are known as follows:

$$\lambda_{ep} = \frac{1.04 + \mu^*\, ln(\Theta_D/1.45T_c)}{(1 - 0.62\mu^*)\, ln(\Theta_D/1.45T_c) - 1.04} \tag{17}$$

We have used the theoretically predicted $T_c$ [26] here. The computed values of $\mu^*$ are listed in **Table 9**. Other estimated parameters related to superconductivity are also given in this table. The calculated $\mu^*$ for $Li_5N$ is relatively low in comparison to the conventional range from 0.10 to 0.20 [114]. This is due to very low TDOS at $E_F$, which ranges from 0.207 (at 150 GPa) to 0.077 (at 350 GPa).

**Table 9**. Calculated TDOS at the Fermi level [($N(E_F)$ in states/eV/f.u.], Debye temperatures ($\Theta_D$ in K), repulsive Coulomb pseudopotential ($\mu^*$), electron-phonon coupling constant ($\lambda_{ep}$), and effective electron–phonon interaction strength ($V_{ep}$ in eV) of $Li_5N$ under pressure in comparison with the available value.

| $P$ (GPa) | $N(E_F)$ | $\Theta_D$ | $\mu^*$ | $\lambda_{ep}$ [This] | $\lambda_{ep}$ [Ref**] | $V_{ep}$ |
|---|---|---|---|---|---|---|
| 150 | 0.207 | 1763.93 | 0.045 | 0.569 | 1.39 | 2.749 |
| 250 | 0.123 | 1985.53 | 0.028 | 0.227 | 0.37 | 1.846 |
| 350 | 0.077 | 2133.92 | 0.019 | 0.118 | 0.19 | 1.532 |

** [26]

As seen in **Table 9**, $\lambda_{ep}$ decreases as 0.569 (150 GPa) → 0.227 (250 GPa) → 0.118 (350 GPa), indicating a transition from moderate to weak electron–phonon coupling under compression. Furthermore, the effective electron–phonon interaction strength $V_{ep}$, estimated from $V_{ep} = \lambda_{ep}/N(E_F)$, decreases as 2.749 eV (150 GPa) < 1.846 eV (250 GPa) < 1.532 eV (350 GPa). This progressive reduction in the electron–phonon interaction is consistent with the strong suppression of $T_c$ with increasing pressure. The pressure-induced weakening of electron–phonon coupling originates from changes in the TDOS near the Fermi level and electron–phonon matrix elements both of which contribute to the magnitude of $\lambda_{ep}$. There is a large discrepancy between the values of $\lambda_{ep}$ obtained in this work and those obtained previously [26]. Wan *et al.* [26] employed Allen-Dynes formalism [115] to calculate $T_c$ where the logarithmically averaged phonon frequency appears as the pre-factor; in MacMillan equation this is replaced by the Debye temperature. For compounds with $\lambda_{ep}$ < 1.5, $T_c$ predicted by McMillan equation is quite accurate [115]. Therefore, the source of disagreement between the two sets of values of $\lambda_{ep}$ is the calculated values of $\Theta_D$ in this work. The Debye temperatures of $Li_5N$ are found to be very high (consistent with its hardness and phonon thermal conductivity) in this study. This results in a much lower $\lambda_{ep}$ in this study for the given value of $T_c$. A self-consistent treatment of superconductivity was beyond the scope of this work, because CASTEP is unable to calculate the Eliashberg spectral function. We intend to do this calculation in future to resolve this issue regarding the values of $\lambda_{ep}$ under pressure for $Li_5Ni$.

### *3.8. Optical properties*

The optical properties are strongly dependent on the electronic features and characterized by the dielectric constant $\varepsilon(\omega)$ which is composed of real [$\varepsilon_1(\omega)$] and imaginary part [$\varepsilon_2(\omega)$] of the dielectric functions. The complex dielectric function describes a material's linear optical response to electromagnetic radiation. In particular, the complex dielectric function $\varepsilon(\omega)$ encompasses contributions from both interband and intraband transitions [116]. The interband dielectric function is obtained using density functional theory. The intraband dielectric function is modeled using the Drude approach, as it plays a significant role in metallic and semi-metallic systems. This intraband response dominates the low-energy optical response, which is governed by parameters such as the plasma frequency and the damping factor; in contrast, the interband electronic transitions primarily examine the high-energy optical response [117].

In this work, all calculations are carried out for the two polarization directions <100> and <001> of the incident photons. Due to the metallic nature of $Li_5N$, a semi-empirical Drude term with a Gaussian smearing of 0.5 eV is incorporated to evaluate the frequency-dependent dielectric constant. Accordingly, a uniform Drude damping parameter of 0.05 eV is applied in all cases, while the plasma frequency is set to 2.0 eV. The optical parameters for different polarization directions indicated the optical anisotropy.

The computed dielectric function of $Li_5N$ under hydrostatic pressures (only shown for 150 GPa and 350 GPa) is illustrated in **Fig. 14**. The negative real part $\varepsilon_1(\omega)$ of the dielectric function, illustrated in **Fig. 14a** for both <100> and <001> directions, clearly demonstrates a Drude-like behavior. The function $\varepsilon_1(\omega)$ in $Li_5N$ is negative within the energy ranges of 0 to 0.96 eV and 0 to 0.98 eV for <001> direction at pressures of 150 GPa and 350 GPa, respectively. For <100> direction, the values are 0 to 0.50 eV and 0 to 0.69 eV at the same pressures. This behavior indicates that $Li_5N$ exhibits typical free electron gas characteristics and metallic properties. Additionally, a strong electromagnetic screening effect prevents electromagnetic waves from propagating, suggesting a high optical reflectivity of the material in these energy ranges [118]. As illustrated in **Fig. 14b**, $\varepsilon_2(\omega)$ of $Li_5N$ demonstrates a pronounced positive value near zero energy, suggests that the material exhibits notable internal absorption and optical attenuation characteristics in the infrared region [118]. Moreover, in the high-energy (ultraviolet) region, both $\varepsilon_2(\omega)$ and $\varepsilon_1(\omega)$ approach zero. Furthermore, in <001> direction, $\varepsilon_2(\omega)$ approaches zero at ~ 0.95 eV for <100> and 1.10 eV at <001> direction, signifying that in that region, the absorption coefficient and reflectivity drop sharply (cf. **Fig. 14 e, g**), and consequently, the loss function (cf. **Fig. 14h**) exhibits a peak [119].

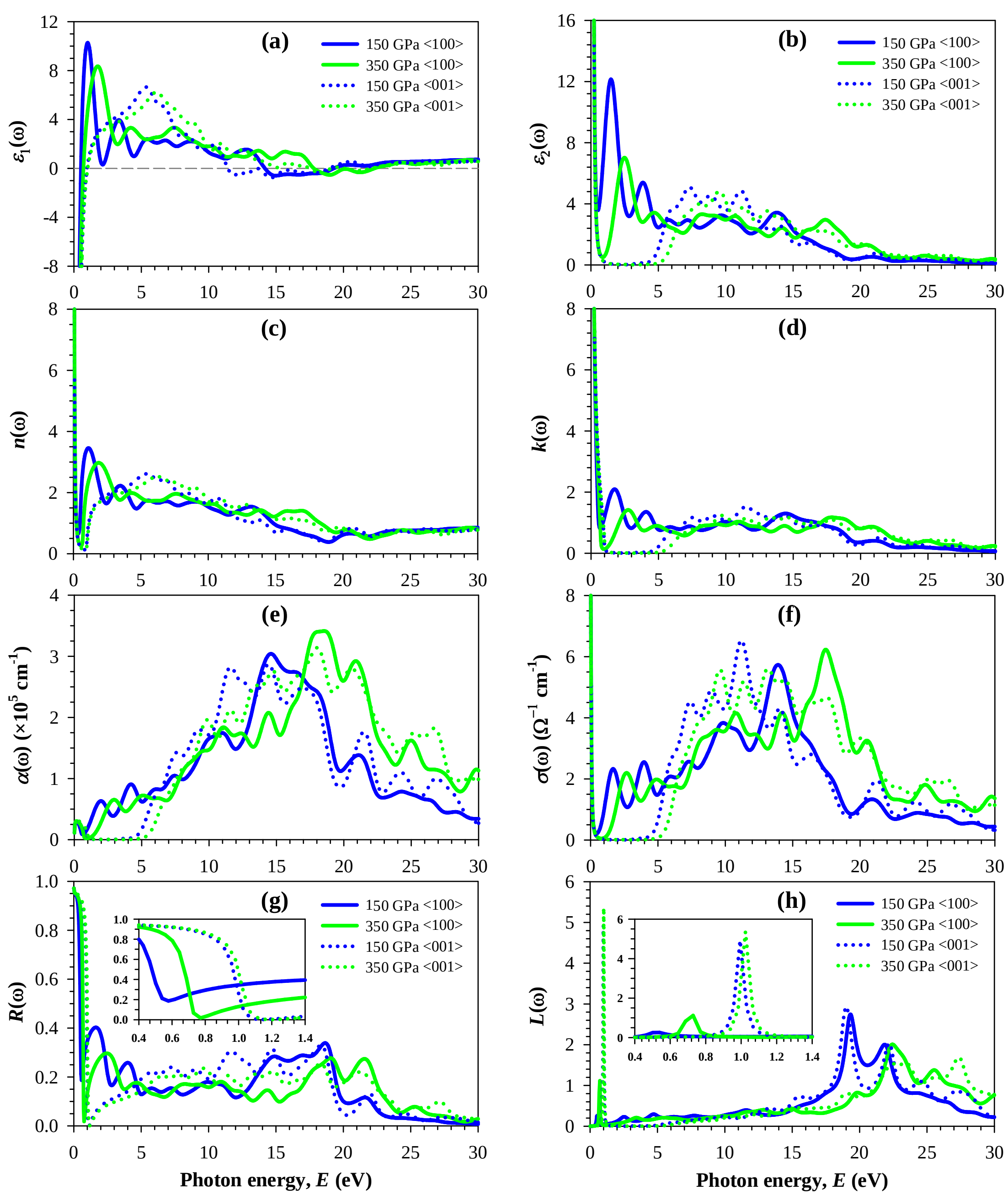

**Fig. 14**. Calculated (a) real ($\varepsilon_1$) and (b) imaginary ($\varepsilon_2$) part of the dielectric function, (c) refractive index ($n$) and (d) extinction coefficient ($k$), (e) absorption coefficient ($\alpha$), (f) conductivity ($\sigma$), (g) reflectivity (inset shows the reflectivity spectra in the low energy region, and (h) loss function (inset shows the pronounced low energy plasma peak) of $Li_5N$ under pressure in <100> and <001> direction.

The changes in refractive index $n(\omega)$ and extinction coefficient $k(\omega)$ with energy are the same as $\varepsilon_1(\omega)$ and $\varepsilon_2(\omega)$, respectively. As shown in **Fig. 14c-d**, both $n(\omega)$ and $k(\omega)$ reach their maximum values at low energies and decrease as the energy increases. This trend is may be due to the declining density of accessible electronic states away from the Fermi level. $Li_5N$ exhibits relatively high refractive indices ($n > 2.0$) in the infrared region across all applied pressures and along both polarization directions, indicating that it is a typical high–refractive-index material with significant potential for applications in infrared photodetection and sensing technologies. The refractive index decreases sharply within the IR region; while it increases again in visible-to-UV region. For <001> direction, a remarkable decrement (almost zero) in $n(\omega)$ is observed at ~ 0.98 eV (350 GPa) in consistent with $\varepsilon_2(\omega)$.

Innately, the shapes of $\alpha(\omega)$ curves are very much similar to the $\sigma(\omega)$ curves in different energies [57]. This behavior can be ascertained by the density of states (DOS) presented in **Fig. 12a-b**. The total DOS at the Fermi level primarily results from the electronic interactions between the Li-2*s* and N-2*p* orbitals that indicates a significant hybridization. Furthermore, the presence of a pseudogap in the TDOS suggests that the bonding states are almost fully occupied by electrons. Consequently, the energy of the incident photon is not sufficient to excite electrons from the valence band to the conduction band. At zero photon energy, $\alpha(\omega)$ hold a threshold value to occur absorption process indicates the metallic nature. The calculated highest absorption peaks arise around the energy range of ~14.5 eV and 18.5 eV for 150 GPa and 350 GPa, respectively, for the <100> polarization (**Fig. 14e**). The peak of $\alpha(\omega)$ arises at ~14 to 17 eV in <001> polarization. **Fig. 14f** illustrates the photoconductivity $\sigma(\omega)$ of $Li_5N$ under pressure. The maximum peak occurs in the <001> direction, reaching 6.53 $\Omega^{-1}cm^{-1}$ at ~11.2 eV under 150 GPa in the UV region, and the highest peak moves toward higher energy when pressure is increased.

The proportion of incident radiation reflected by the material can be identified by reflectivity $R(\omega)$. The finite values of reflectivity at zero energy ($\omega = 0$) are referred to as the static values of reflectivity, denoted as $R(0)$. The calculated values of $R(0)$ are approximately 0.97 within the whole pressure range (**Fig. 14g**). As the energy rises, the reflectivity starts to decrease. $R(\omega)$ decreases to minima of ~0.001 at 1.14 eV and 0.0008 at 1.19 eV for pressures of 150 GPa and 350 GPa, respectively, in the <001> direction. The reflection spectra drops drastically at the end of IR range indicates that the compound under study is expected to have high transmittance in this region [120]. The inset of **Fig. 14g** clearly illustrates this trend in the IR region. In the energy range of 5–15 eV, $R(\omega)$ undergoes successive fluctuations, with its value consistently below 0.3. Then, around 17 eV, $R(\omega)$ reaches maxima of ~40% and 33% for <100> and <001> polarizations, respectively at 150 GPa. After ~18 eV, it decreases again, approaching 0.05 in <001> direction. Afterward, $R(\omega)$ begins to increase once more, reaches peaks at 21.5–22.5 eV, and finally rapidly drops with minor fluctuations to zero at the end of ~30 eV. Similar behavior is observed for 350 GPa with the peaks shifting at higher energies.

The energy loss function $L(\omega)$ is widely used to describe the energy dissipation of fast-moving electrons as they traverse a solid. A characteristic correlation exists between the reflectivity spectrum, $R(\omega)$, and the loss function, $L(\omega)$, where the peak position of $L(\omega)$ generally appears at the descending edge of $R(\omega)$ [57]. As seen in **Fig. 14h**, the first $L(\omega)$ peak appears in the IR region. The inset of **Fig. 14h** clearly illustrates the trend of the apex peaks in the IR region, with the peaks shifting towards higher energy when pressure is raised from 150 GPa to 350 GPa.

Overall, the calculated optical spectra of $Li_5N$ electride exhibit pronounced anisotropy and a broadening at higher pressures. This anisotropy is particularly pronounced in the high-pressure phase of the electride, where continuous modification of the microscopic structure by external pressure leads to notable variations in dielectric response [121]. $Li_5N$ might be used as an excellent reflector in the infrared region. High refractive index in the visible region at 350 GPa and high absorption coefficient in the ultraviolet regime also indicate the potential of $Li_5N$ in optoelectronic applications.

## 4. Conclusion

High pressure (150-350 GPa) hexagonal $Li_5N$ have been investigated successfully using *ab-initio* study through density functional theory. To explore the thermodynamic stability of $Li_5N$, its formation ($E_f$) and cohesive ($E_c$) energies are calculated. The mechanical stability of $Li_5N$ is further examined using elastic constants by Born-Huang stability criteria, and found to be stable within the studied range of pressure. Phonon dispersion curves clearly demonstrate dynamic stability at pressures ranging from 100 GPa to 382 GPa. However, at pressures below 100 GPa and beyond 382 GPa, the system becomes dynamically unstable.

Mechanical properties are evaluated using elastic constants. Estimated Poisson's ratio ($\sigma$) indicates the bonding forces are central as $\sigma$ lies within 0.29~0.34. Moreover, $Li_5N$ exhibits ductile nature in view of estimated Poisson's ratio, Pugh's ratio ($G/B$, lies within 0.49~0.37) and Cauchy pressure ($P_C > 0$). In view of various anisotropy indices, $Li_5N$ exhibits anisotropic nature and the increment of pressure from 150 GPa enhances its anisotropy with few anomalies. The combination of high hardness and pressure-induced strengthening makes this compound a promising candidate for applications in high-pressure environments. The acoustic velocities, Debye and melting temperatures are found to be dependent on pressure; all intensified as the pressure rises from 150 GPa to 350 GPa. $Li_5N$ is expected to be thermally conductive for having high Debye temperatures ranging from ~1764 K to ~2134 K. The melting temperature of the compound is extremely high. The metallic conductivity of this compound within 150-350 GPa pressure is predicted from electronic structure analysis. Both ionic and covalent bonding are present in $Li_5N$ in view of the Mulliken atomic population analysis and charge density mapping. The ionic bonding weakens at higher pressures, while covalency increases. The optical response exhibits metallic features, including dominant intraband contributions at low energies, strong ultraviolet absorption (~ $10^5$ $cm^{-1}$), high low-energy reflectivity (~0.97), and plasma resonance behavior.

The pressure dependent superconducting features of $Li_5N$ are investigated. This study implies that very high Debye temperature facilitates high-temperature superconductivity in this compound. As pressure increases, the density of states at the Fermi level and the electron-phonon interaction energy decreases which has a strong detrimental effect on superconducting transition temperature.

To summarize, this study investigates the physical properties of the high-pressure hexagonal phase of $Li_5N$ electride. Most of these properties were unexplored and we hope the results presented herein would provide guidelines for future investigations.

**Acknowledgements**

M.A.H.S. acknowledges the fellowship from the Science and Technology Fellowship Trust, Ministry of Science and Technology, Bangladesh for his Ph.D. research. S.H.N. acknowledges the research grant (1151/5/52/RU/Science-07/19-20) from the Faculty of Science, University of Rajshahi, Bangladesh, which partly supported this work.

**Data availability**

Data will be made available from the corresponding author on reasonable request.

**Declaration of competing interest**

The authors declare that they have no known competing financial interests or personal relationships that could have appeared to influence the work reported in this paper.

**CRediT authorship contribution statement**

**M.A.H. Shah**: Investigation, Methodology, Data curation, Visualization, Formal analysis, Writing–original draft, **S.H. Naqib**: Conceptualization, Validation, Supervision, Resources, Project administration, Writing-review & editing.